\documentclass[journal]{IEEEtran}
\ifCLASSINFOpdf
\else
\fi
\usepackage{amsmath}
\usepackage{amssymb}

\usepackage{amsfonts}
\usepackage[nice]{nicefrac}
\usepackage[pdftex]{graphicx}
\usepackage{array}
\usepackage[caption=false,font=footnotesize]{subfig}
\usepackage{bm}
\usepackage{stfloats}
\usepackage{color}
\usepackage{cite}
\usepackage{url}
\begin{document}
%
\title{Robust MIMO Radar Waveform-Filter Design for Extended Target Detection in the Presence of Multipath}

%
%
%

\author{Zhou~Xu,
        Chongyi~Fan,
        Jian~Wang,
        and~Xiaotao~Huang,~\IEEEmembership{Senior Member,~IEEE}
\thanks{Manuscript received *** **, 2018; accepted *** **, 2018. Date of publication *** **, 2018; date of current version *** **, 2018.\emph{(Corresponding author: Chongyi Fan.)}}
\thanks{Z. Xu is with the College of Electronic Science and
	Engineering, National University of Defense Technology, Changsha 410073, and also with the Colledge of Electronic Countermeasure,National University of Defense Technology, Hefei 230037,	China (e-mail: zhouzhou900521@126.com).}
\thanks{C. Fan, J. Wang and X. Huang are with the College of Electronic Science and Engineering,
	National University of Defense Technology, Changsha 410073, China(e-mail: chongyifan@nudt.edu.cn).}
\thanks{Color versions of one or more of the figures in this paper are available
	online at http://ieeexplore.ieee.org.}
\thanks{Digital Object Identifier **.****/TIP.2018.******.}}

%
%

\markboth{IEEE TRANSACTIONS ON SIGNAL PROCESSING,~Vol.~**, No.~**, **~2020}
{Xu \MakeLowercase{\textit{et al.}}:ROBUST MIMO RADAR WAVEFORM DESIGN FOR EXTENDED TARGET DETECTION IN THE PRESENCE OF MULTIPATH}
%




\maketitle

\begin{abstract}
The existence of multipath brings extra "looks" of targets. This paper considers the extended target detection problem with a narrow band Multiple-Input Multiple-Output(MIMO) radar in the presence of multipath from the view of waveform-filter design. The goal is to maximize the worst-case Signal-to-Interference-pulse-Noise Ratio(SINR) at the receiver against the uncertainties of the target and multipath reflection coefficients. Moreover, a Constant Modulus Constraint(CMC) is imposed on the transmit waveform to meet the actual demands of radar. Two types of uncertainty sets are taken into consideration. One is the spherical uncertainty set. In this case, the max-min waveform-filter design problem belongs to the non-convex concave minimax problems, and the inner minimization problem is converted to a maximization problem based on Lagrange duality with the strong duality property. Then the optimal waveform is optimized with Semi-Definite Relaxation(SDR) and randomization schemes. Therefore, we call the optimization algorithm Duality Maximization Semi-Definite Relaxation(DMSDR). Additionally, we further study the case of annular uncertainty set which belongs to non-convex non-concave minimax problems. In order to address it, the SDR is utilized to approximate the inner minimization problem with a convex problem, then the inner minimization problem is reformulated as a maximization problem based on Lagrange duality. We resort to a sequential optimization procedure alternating between two SDR problems to optimize the covariance matrix of transmit waveform and receive filter, so we call the algorithm Duality Maximization Double Semi-Definite Relaxation(DMDSDR). The convergences of DMDSDR are proved theoretically. Finally, numerical results highlight the
effectiveness and competitiveness of the proposed algorithms as well as the optimized waveform-filter pair.   
\end{abstract}

\begin{IEEEkeywords}
Extended target detection, Waveform-filter design, Worst-case SINR, Minimax, Lagrange duality, Semi-Definite Relaxation(SDR).  
\end{IEEEkeywords}

%
\IEEEpeerreviewmaketitle

\section{Introduction}
%
%
%
%

\IEEEPARstart{R}{ecently}, Multiple-Input Multiple-Output(MIMO) radar has draw increasing attention from researchers for its excellent parameter identifiability and waveform diversity{\cite{LiMIMO,LiMIMO2008,LiOnparameter,bliss2003multiple}, which can significantly improve the target detection performance. Different from the point-like targets, extended radar targets exhibit complicated scattering characteristics instead of a scaled
and attenuated version of the transmit waveform back to the radar receiver. Considering extended targets are more common in practical radar detection, many researchers have investigated the extended target detection and recognition problem via waveform optimization{\cite{bell1993information,yang2007mimo,goodman2007adaptive,meng2012radar,jiu2012minimax,yang2007minimax,chen2009mimo,karbasi2015robust}}. 

In most cases, the scattering behavior
of the extended target is characterized by the Target Impulse Response(TIR) or Power Spectral Density(PSD){\cite{kay2007optimal,li1996scattering,bell1993information,jiu2012minimax,yang2007mimo,yang2007minimax,karbasi2015robust}}. In {\cite{bell1993information}}, two kinds of waveform are studied. The first one is the optimal detection waveform, where the maximum  Signal-to-Interference pulse Noise Ratio(SINR) criterion is used with a deterministic TIR of the extended target, since the detection probability monotonically increases with respect to SINR{\cite{DeMaioCode}}; The second one is the optimal estimation waveform, where the extended target scattering characteristics are modeled as a Gaussian random process with a deterministic Power Spectral Density(PSD), and the maximum mutual information criterion is used for target identification and classification. Yang \textit{et.al.} generalize the waveform design problem to the MIMO case{\cite{yang2007mimo}}, where both the Mean-Square Error(MSE) and MI criteria are studied, and results show that the two criteria lead to the same optimal waveform when the energy constraint is imposed on the transmitter.

However, the TIR or PSD information of target can't be accurately obtained in practice due to a variety of factors{\cite{kay2007optimal,li1996scattering}}. Therefore, robust methods, as widely used in other applications, should be taken into consideration {\cite{kassam1985robust,ben2002robust,kim2006robust,li2004doubly,NaghshADoppler,wang2017robust,jin2019local,razaviyayn2020nonconvex,wang2020improved}}. In {\cite{yang2007minimax}}, the authors extend their work in {\cite{yang2007mimo}} to deal with the uncertain PSD, and design waveform based on the worst-case performance. Aiming at maximizing the worst-case SINR, authors in\cite{wang2017robust} consider the joint design of MIMO radar waveform and filter with imprecise knowledge about target doppler and direction, where the Lagrange duality method is developed to deal with the inner minimization problem. As to the extended target detection, the spherical uncertainty set is usually used to describe the imprecise knowledge of TIR{\cite{chen2009mimo,karbasi2015robust}}. In\cite{chen2009mimo}, an iterative algorithm is proposed for joint design of transmit waveform and receive filter to deal with the spherical uncertainty of TIR. At each iteration, the algorithm maximize the worst SINR by solving a minimax problem and get a monotonically increased worst SINR. However, only the energy constraint on the transmit waveform is considered in {\cite{chen2009mimo}}, which might lead to the undesired waveform with a high Peak-to-Average Ratio(PAR){\cite{skolnik1990radar,patton2009autocorrelation}}. In {\cite{karbasi2015robust}}, a PAR constraint is imposed on the transmit waveform to meet the actual demands of radar transmitter and two types of uncertainty sets are studied. The first one is that the TIRs are possibly chosen from a finite set. The second one is the case of spherical uncertainty set over a prescribed TIR. Different from the algorithm in {\cite{chen2009mimo}}, the authors in {\cite{karbasi2015robust}} randomly pick several samples to approximate the spherical uncertainty set, then the waveform covariance and the filter covariance are alternatively optimized by two Semi-Definite Programming(SDP). However, as we will see later, insufficient samples might lead to SINR losses when constructing the spherical uncertainty set.

Multipath is very common in radar detection. Though the existence of multipath challenges radar detection and recognition, it also increases the spatial diversity of the radar system by providing extra “looks” at the target and improves 
radar detection and recognition performance{\cite{mecca2006mimo,fishler2006spatial,sen2010ofdm,SenAdaptive,SenAdaptiveOFDM}}.  
In {\cite{sen2010ofdm,SenAdaptive,SenAdaptiveOFDM}}, waveform design schemes are proposed for target detection and tracking in the presence of multipath. However, only the point-like targets are studied in the previously mentioned literatures, and the authors don't take the robustness into consideration. 

In this paper, robust waveform-filter design for the extended target detection with a narrow band radar is considered by exploiting the spatial diversity provided by multipath. The knowledge of scattering coefficients is assumed to be imprecise, and two types of uncertainty sets are studied, namely, the spherical uncertainty set and the annular uncertainty set. The spherical uncertainty set is used to describe the imprecise knowledge of a prescribed TIR, and the larger sphere radius implies more inaccurate prescribed TIR. The Duality Maximization Semi-Definite Relaxation(DMSDR) algorithm is devised to solve the worst-case SINR optimization problem under the spherical uncertainty set. Meanwhile, the annular uncertainty set is used to model the random phase of scattering, which means that the scattering amplitude of target or scatterers can be roughly estimated, but the scattering phase is totally random. The Duality Maximization Double Semi-Definite Relaxation(DMDSDR) algorithm is devised to solve the worst-case SINR optimization problem under the annular uncertainty set. It is worth pointing out that although the two proposed algorithms are devised for the narrow band MIMO radar in the presence of multipath, they can also be applied to extended target detection with a high resolution MIMO radar in settings without multipath, because the structure of optimization problems is almost the same. Specially, our work makes the
following contributions:

1) \textit{Robust MIMO waveform-filter design against two types of uncertainty sets}: Based on the worst-case SINR criterion, joint transmit waveform and receive filter design under the spherical uncertainty set as well as the annular uncertainty set are studied. Different from the spherical uncertainty set, the annular uncertainty set describes the totally random phase of the scatterers and is non-convex. To our best knowledge, there is rarely literature discussing the annular uncertainty set.

2) \textit{The DMSDR algorithm to solve the spherical uncertainty set problem}: The robust waveform-filter design problem against the spherical uncertainty set belongs to the non-convex concave minimax problems. Different from the sequential optimization procedure in   {\cite{chen2009mimo,wang2017robust,karbasi2015robust}}, we prove that the optimal filter can be analytically calculated first. Then, the robust waveform-filter design problem is converted to a maximization problem by duality theory{\cite{BoydConvex}}. Thus, the waveform covariance can be optimized through Semi-Definite Relaxation(SDR){\cite{LuoSemidefinite}} without sequential optimization.

3) \textit{The DMDSDR algorithm to solve the annular uncertainty set problem}: The robust waveform-filter design problem against the annular uncertainty set belongs to the non-convex non-concave minimax problems owing to the non-convex annular uncertainty set. In order to address the problem, the inner minimization problem is approximated by a convex problem with SDR whose duality problem is derived in the paper. Therefore, the robust waveform-filter design problem can be converted into a maximization problem through duality theory. The transmit waveform covariance and the receive filter covariance are alternatively optimized with SDR. Moreover, we theoretically proved that the proposed DMDSDR converges to a stationary point.

4) \textit{Analyses and experiments for DMSDR and DMDDSR}: The computational complexities of DMSDR and DMDSDR are analyzed. And the numerical experiments are carried out to verify the effectiveness of the proposed algorithms. The results highlight the robustness of the output SINR. 
 
The remainder of this paper is organized as follows. Section II formulates the robust  waveform-filter design with the narrow band MIMO radar in multipath scenario, and builds up its signal model. Section III introduces the DMSDR algorithm against the spherical uncertainty set. In addition, the computational complexities are analyzed. Section IV introduces the DMDSDR algorithm against the annular uncertainty set, and gives a further discussion on its convergences and computational complexities. Section V provides several numerical experiments to demonstrate the effectiveness of the proposed algorithms, and exhibits the performance of the waveform-filter pair. Finally, Section V draws conclusions.

\textit{Notations:} Throughout this paper, scalars are denoted by italic letters(e.g., \textit{a}, \textit{A}); vectors are denoted by bold italic lowercase letters(e.g., $\bm{a}$) and $\bm{a}(i)$ denotes the $i$th element of $\bm{a}$; ${\bm{e}}_i$ denotes the unit vector with the $i$th element being 1; matrices are denoted by bold italic capital letters(e.g., $\bm{A}$) and $\bm{A}(i,j)$ denotes the element in the $i$th row and $j$th column of $\bm{A}$; ${\bf{I}}_N$ is the unit matrix with size $N$. Superscript ${\left( \cdot\right) ^{\rm{T}}}$ and ${\left( \cdot\right) ^{\rm{H}}}$ denote transpose and conjugate transpose, respectively. tr($\cdot$) denotes the trace of a square matrix. vec($\cdot$) denotes the operator of column-wise stacking a matrix. $\otimes$ and $\odot$ represent Kronecker product and Hadamard product, respectively. diag($\bm{a}$) denotes the diagonal matrix with the diagonal elements formed by $\bm{a}$, while diag($\bm{A}$) denotes the vector with elements formed by the diagonal elements of $\bm{A}$. $\left\lceil {a} \right\rceil $ denotes the maximum integer less than $a$; arg($\bm{a}$) denotes the vector consisting of the phase angle of $\bm{a}(i)$ and $\left| \bm{a}\right| $ denotes the vector consisting of the modulus of ${\bm{a}(i)}$. The representation $\bm{A} \succ 0(\bm{A}\succeq 0)$ means $\bm{A}$ is positive definite(semi-definite). 
\section{Problem Formulation and Signal Model}
A narrow band MIMO radar is used to detect an extended target, and we assume that there are some strong scatterers in the scenario, which provide extra "looks"(multipath) of the target from different angles. The scenario is depicted in Fig.\ref{Scenario}, where the red arrows(i.e. $\overrightarrow{\rm{OA}}$, $\overrightarrow{\rm{OA}}-\overrightarrow{\rm{AT}}$) and blue arrows(i.e. $\overrightarrow{\rm{AO}}$, $\overrightarrow{\rm{TA}}-\overrightarrow{\rm{AO}}$) represent the possible transmit and receive paths, respectively. 

%
%
 
\begin{figure}[!t]
  	\centering
 	\includegraphics[width=3.5in,height=2.6in]{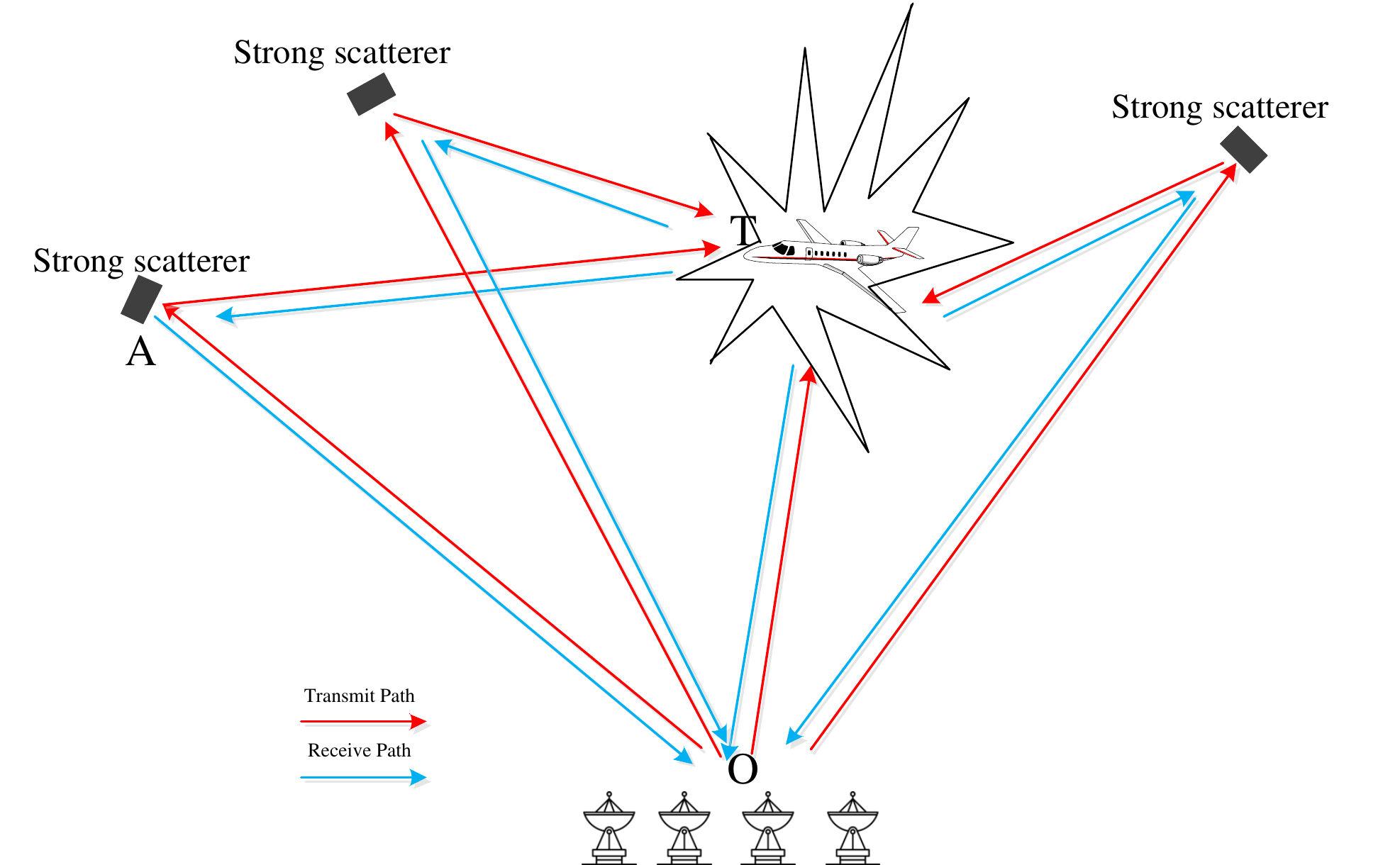}
  \caption{Extended target detection diagram}
  \label{Scenario}
\end{figure}
\subsection{Problem Formulation}
As described in Fig.\ref{Scenario}, the radar is located at the scenario center O detecting the moving extended target T. Different from the conventional scenarios, some strong scatterers that might cause the multipath returns exist in the scenario. According to the transmit and receive paths, the received signal can be formulated as
\begin{equation}
{\bm{y}(\bm{s})} = {\alpha _0}{{\bm{y}}_0(\bm{s})} + \sum\limits_{k = 1}^K {{\rho _k}{\alpha _k}{{{\bm{y}}_k(\bm{s})}}}  + {\bm{n}},
\label{SignalModelEq}
\end{equation}          
where ${{\bm{y}}_0(\bm{s})}$ and ${{{\bm{y}}}_k}(k\ge 1)$ denote the direct returns and the {\textit{k}}th multipath returns from the target, respectively, both of which depend on the transmit waveform $\bm{s}$. ${\bm{n}}$ denotes interference and noise. ${\alpha _k} (k = 0,1,2,...,K)$ is the target complex scattering coefficient of the {\textit{k}}th propagation path, and ${\rho _k} (k = 1,2,...,K)$ is the complex attenuation coefficient due to multipath propagation. In order to simplify the notations being used, we use ${\bm{y}_k}(k=0,1,...,K)$ instead of ${\bm{y}_k(\bm{s})}(k=0,1,...,K)$ in the followings if no confusion wil be caused.

In order to achieve the best detection performance, we want to maximize the output SINR by jointly designing transmit waveform and receive filter with the multipath information also being utilized. Let \textit{\textbf{w}} be the filter vector, the output SINR at the receiver can be expressed as 
\begin{equation} 
{\rm{SINR}}({\bm{w}},{\bm{s}}) = \frac{{{{\left| {{{\bm{w}}^{\rm{H}}}\left( {{\alpha _0}{{\bm{y}}_0} + \sum\limits_{k = 1}^K {{\rho _k}{\alpha _k}{{{\bm{y}}}_k}} } \right)} \right|}^2}}}{{{{\bm{w}}^{\rm{H}}}{{\bm{R}}_n}{\bm{w}}}},
\label{SINROptEq}
\end{equation}
where ${\bm{s}}$ is the transmit waveform and ${{\bm{R}}_{n}}$ denotes the covariance matrix of interference and noise. 

Unfortunately, ${\alpha _k}$ is so sensitive with respect to the small changes in target orientation that it is impossible to get the precise knowledge of it\cite{karbasi2015robust,1992Variability}. In addition, multipath attenuation coefficient depends on a variety of factors, for instance, the scatter material, the frequency of incident signals, which make it impossible to get the precise knowledge of ${\rho _k}$. Inspired by the robust idea, we consider the robust(worst-case) SINR criterion with some partial knowledge for $\bm{\alpha}$ and $\bm{\rho}$. To this end, we formulate the robust waveform-filter design problem as 
\begin{equation}
\begin{array}{l}
\mathop {{\rm{max}}}\limits_{{\bm{w}},{\bm{s}}} \mathop {{\rm{ min}}}\limits_{{\bm{\alpha }},{\bm{\rho }}} {\rm{ SINR}}({\bm{w}},{\bm{s}},{\bm{\alpha }},{\bm{\rho }})\\
s.t.\quad \left( {{\bm{\alpha }},{\bm{\rho }}} \right) \in \Theta ,\left( {{\bm{w}},{\bm{s}}} \right) \in \Upsilon 
\end{array}, 
\label{RobustSINROptEq} 
\end{equation}
where ${\bm{\alpha }} = {[{\alpha _0},{\alpha _1},...,{\alpha _K}]^{\rm{T}}} \in {\mathbb{C}}^{{K+1}}$, ${\bm{\rho }} = {[{1},{\rho _1},...,{\rho _K}]^{\rm{T}}}\in {\mathbb{C}}^{{K+1}}$, $\Theta $ denotes the uncertainty set to which  $\left({\bm{\alpha}},{\bm{\rho}} \right)$ belongs. And $\Upsilon $ denotes the constraint set on $\left({\bm{w}},{\bm{s}} \right)$. Substitute (\ref{SINROptEq}) into (\ref{RobustSINROptEq}) by introducing an auxiliary variable $\bm{u}=\bm{\alpha}\odot\bm{\rho} \in {\mathbb{C}}^{{K+1}}$, then we reformulate  (\ref{RobustSINROptEq}) as a more compact form
\begin{equation}
\begin{array}{l}
\mathop {{\rm{max}}}\limits_{{\bm{w}},{\bm{s}}} \mathop {{\rm{ min}}}\limits_{\bm{u}} {\rm{ }}\frac{{{{\left| {{{\bm{w}}^{\rm{H}}}{\bm{Y(s)u}}} \right|}^2}}}{{{{\bm{w}}^{\rm{H}}}{{\bm{R}}_n}{\bm{w}}}}\\
s.t.\quad {\bm{u}} \in \Theta ,\left( {{\bm{w}},{\bm{s}}} \right) \in \Upsilon 
\end{array},
\end{equation}
where ${\bm{Y(s)}} = [{{\bm{y}}_0},{{\bm{y}}_1},{{\bm{ y}}_2},...,{{\bm{ y}}_K}]$. In order to simplify the notations being used, we use ${\bm{Y}}$ instead of ${\bm{Y(s)}}$ in the followings if no confusion wil be caused.

In our paper, two types of uncertainty sets are studied. The first one is the spherical uncertainty, i.e. $\bm{u}$ belongs to a scaled ball centered around an a priori known $\bm{u_0}$. The second one is the annular uncertainty, i.e. the amplitude of $\bm{u}$ can be estimated roughly in advance, but the phase is totally random. It is worth noting that, the two types of uncertainty sets are essentially different, because the spherical uncertainty set is convex, while the annular uncertainty set is non-convex. The geometric forms of two uncertainty sets are briefly outlined in Fig.\ref{Uncertainty set}
for a more intuitive understanding. 
 
As to the constraint set $\Upsilon$, the Constant Modulus Constraint(CMC) on the transmit waveform is considered. Thus, the following max-min problems ${{\cal P}_1}$ and ${\tilde{\cal P}_1}$ are studied,

\begin{equation}
{{\cal P}_1}\left\{ \begin{array}{l}
\mathop {{\rm{max}}}\limits_{{\bm{w}},{\bm{s}}} \mathop {{\rm{ min}}}\limits_{\bm{u}} {\rm{ }}\frac{{{{\left| {{{\bm{w}}^{\rm{H}}}{\bm{Yu}}} \right|}^2}}}{{{{\bm{w}}^{\rm{H}}}{{\bm{R}}_n}{\bm{w}}}}\\
s.t.\quad {\left\| {{\bm{u}} - {{\bm{u}}_0}} \right\|_2} \le r,\left| {{\bm{s}(i)}} \right| = 1
\end{array} \right.,
\label{RobustSINRType1}
\end{equation}

\begin{equation}
{\tilde{\cal P}_1}\left\{ \begin{array}{l}
\mathop {{\rm{max}}}\limits_{{\bm{w}},{\bm{s}}} \mathop {{\rm{ min}}}\limits_{\bm{u}} {\rm{ }}\frac{{{{\left| {{{\bm{w}}^{\rm{H}}}{\bm{Yu}}} \right|}^2}}}{{{{\bm{w}}^{\rm{H}}}{{\bm{R}}_n}{\bm{w}}}}\\
s.t.\quad {\bm{\eta}(k)} \le \left| {{\bm{u}(k)}} \right| \le {\bm{\xi}(k)},\left| {\bm{s}(i)} \right| = 1
\end{array} \right. ,
\label{RobustSINRType2}
\end{equation}
where ${\bm{\eta}(k)}$ and ${\bm{\xi}(k)}$ denote the lower bound and upper bound of scattering amplitude, respectively. 
 
\begin{figure}[!t]
	\centering
	\begin{tabular}{c}
		\includegraphics[width=2.2in,height=2.8in]{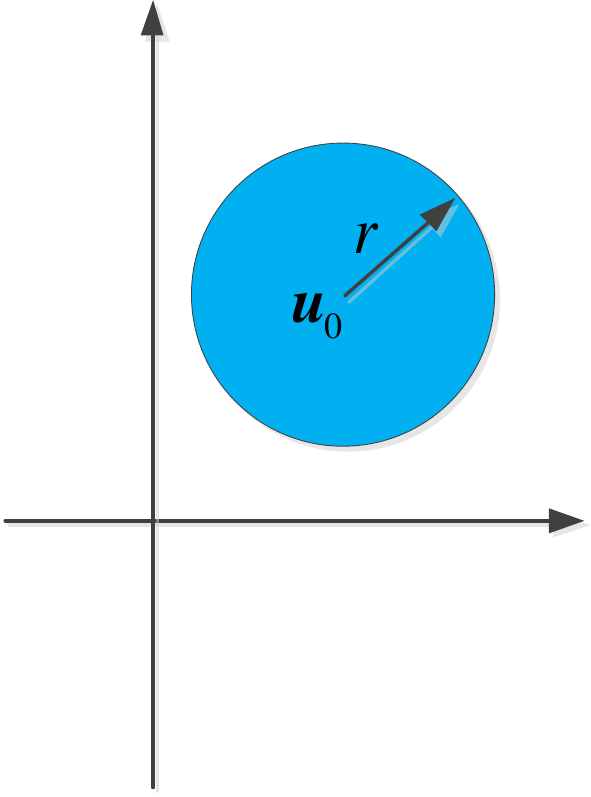}\\
		(a)\\
		\includegraphics[width=2.2in,height=2.6in]{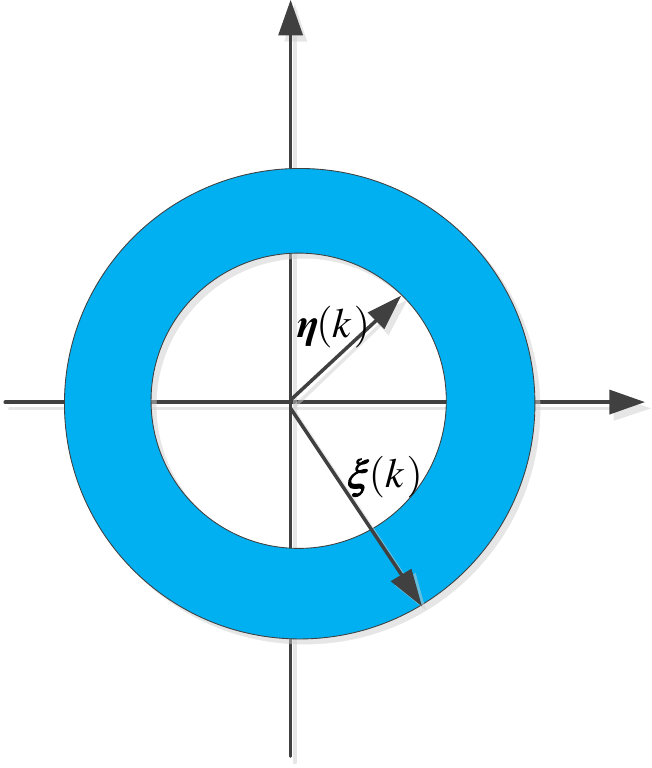}\\
		(b)\\
	\end{tabular}
	\centering
	\caption{Two types of uncertainty sets: (a) The spherical uncertainty set. (b) The annular uncertainty set.}
	\label{Uncertainty set}
\end{figure}

\subsection{Signal model of $\bm{y}_0$}
Consider the target detection problem for a colocated MIMO radar with ${N_T}$ transmitters and ${N_R}$ receivers in the multipath scenario described in Fig.\ref{Scenario}. The waveform transmitted by the \textit{n}th transmitter with \textit{L} samples is denoted by ${{\bm{s}}_n} = {[{{\bm{s}}_n}(1),{{\bm{s}}_n}(2),...,{\bm{s}}_n(L)]^{\rm{T}}}$, then the transmitting matrix for MIMO can be represented as ${\bm{S}} = {[{{\bm{s}}_1},{{\bm{s}}_2},...,{{\bm{s}}_{{N_T}}}]^{\rm{T}}} \in {\mathbb{C}}^{{N_T} \times L}$. Regardless of signal amplitude, the baseband signal of direct path returns from target can be formulated as
\begin{equation}
{{\bm{Y}}_{0}} = {\bm{b}}({\theta _{0}}){{\bm{a}}^{\rm{T}}}({\theta _{0}}){\bm{S}}{\bm{J}_0},
\label{DPTargetEq}
\end{equation}  
where ${\bm{b}}({\theta _{0}})$ and ${\bm{a}}({\theta _{0}})$ denote the transmit steering vector and receive steering vector at target line of sight ${\theta _{0}}$, respectively. Denoted  ${\bm{J}_l} \in {\mathbb{C}}^{{L} \times P}$ by the shift matrix 
\begin{equation}
	{\bm{J}_l}(i,j) = \left\{ \begin{array}{l}
	1{\kern 1pt} ,\;{\rm{if}}\;i - j + l = 0\\
	0,\;{\rm{if}}\;i - j + l \ne 0
	\end{array} \right..
\end{equation}
Note that the multipath propagation distance is always longer than the distance of direct path. In order to record all samples of returns in fast time, the column number of ${\bm{J}_l}$ must be larger than $L$, namely $P>L$.

Let ${{\bm{y}}_{0}} = {\rm{vec}}\left( {{{\bm{Y}}_{0}}} \right)$ and ${\bm{s}} = {\rm{vec}}({\bm{S}})$ , then we have
\begin{equation}
\begin{array}{l}
{{\bm{y}}_{0}} = {\rm{vec}}\left( {{{\bm{Y}}_{0}}} \right)\\
{\rm{ \quad \ =  }}\left( {{{{\bm{J}_0^{\rm{T}}}}} \otimes \left( { {\bm{b}}({\theta _{0}}){{\bm{a}}^{\rm{T}}}({\theta _{0}})} \right)}\right) {\bm{s}}\\
{\rm{ \quad \ =  }}{\bm{A}_0}{\bm{s}}
\end{array}.
\label{DPTargetReceiveEq}
\end{equation}
\subsection{Signal model of $\bm{{y}}_k(k\ge 1)$}
As to $\bm{{y}}_k$, we only consider the first order multipath returns(see Fig.\ref{Scenario}) for simplicity, which means that the energy of the second or higher order multipath is small enough to be neglected. We category the multipath signal into two groups. The first one is that the transmitted signal reaches the target with once reflection, and is received from the line of \vspace{3pt} sight(see $\overrightarrow{\rm{OA}}$-$\overrightarrow{\rm{AT}}$-$\overrightarrow{\rm{TO}}$ in Fig.\ref{Scenario}). The second one is that the transmitted signal reaches the target by the line of sight\vspace{3pt}, and is received from the direction of multipath reflection(see $\overrightarrow{\rm{OT}}$-$\overrightarrow{\rm{TA}}$-$\overrightarrow{\rm{AO}}$ in Fig.\ref{Scenario}). 

For the first group of multipath signal, we slightly modify (\ref{DPTargetEq}) and get the baseband signal of multipath returns, formulating as
\begin{equation}
{{{\bm{Y}}}_{k}} = {\bm{b}}({\theta _{0}}){{\bm{a}}^{\rm{T}}}({\theta _{k}}){\bm{S}}{\bm{J}_{l_k}},
\label{MPTargetEqFirstType}
\end{equation}   
where ${\theta _{k}}$ denotes the direction of a certain multipath, and ${l_k} \geq 0$ denotes the relative delay in the fast time domain.
According to the derivation of (\ref{DPTargetEq}) and (\ref{DPTargetReceiveEq}), the multipath signal model ${{{\bm{y}}}_{k}}$ in the first group is given by 
\begin{equation}
\begin{array}{l}
{{{\bm{y}}}_{k}} =\left( { {{{\bm{J}_{l_k}^{\rm{T}}}}} \otimes \left( { {\bm{b}}({\theta _{k}}){{\bm{a}}^{\rm{T}}}({\theta _{0}})} \right)}\right) {\bm{s}}
\end{array}.
\label{MPTargetReceiveEqFirstType}
\end{equation}

It is worth pointing out that the multipath returns in the second group have the same time delay as its counterpart in the first group due to the same transmit-receive path. Similarly, multipath signal model ${{{\bm{y}}}_{k}}$ in the second group is given by 
\begin{equation}
\begin{array}{l}
{{{\bm{y}}}_{k}} =\left( { {{{\bm{J}_{l_k}^{\rm{T}}}}} \otimes \left( { {\bm{b}}({\theta _{0}}){{\bm{a}}^{\rm{T}}}({\theta _{k}})} \right)}\right) {\bm{s}}
\end{array}.
\label{MPTargetReceiveEqSecondType}
\end{equation}

Combing the two groups together, we formulate ${{{\bm{y}}}_{k}}$ as 

\begin{equation}
{{\bm{ y}}_k} = \left\{ \begin{array}{l}
\left( { {\bm{J}}_{{l_m}}^{\rm{T}} \otimes {\bm{b}}\left( {{\theta _0}} \right){{\bm{a}}^{\rm{T}}}\left( {{\theta _m}} \right)} \right){\bm{s}},\;k \; \rm{is \; odd}\vspace{3pt}\\ 
\left( { {\bm{J}}_{{l_m}}^{\rm{T}} \otimes {\bm{b}}\left( {{\theta _m}} \right){{\bm{a}}^{\rm{T}}}\left( {{\theta _0}} \right)} \right){\bm{s}}\;, k \; \rm{is \; even}
\end{array} \right.,
\label{MPTargetReceiveEq}
\end{equation}
where $m=1,2,..., \left\lceil {\frac{K}{2}} \right\rceil $ denotes the number of strong scatterers in the scenario. Moreover, if the scattering reciprocities hold for the target and scatterers\cite{tsang2004scattering}, namely ${\rho _{2m - 1}}{\alpha _{2m - 1}} = {\rho _{2m}}{\alpha _{2m}}$, ${{\bm{ y}}_k}$ can be simplified as
\begin{equation}
\begin{array}{l}
{{{\bm{ y}}}_k} = \left( { {\bm{J}}_{{l_k}}^{\rm{T}} \otimes \left( {{\bm{b}}\left( {{\theta _0}} \right){{\bm{a}}^{\rm{T}}}\left( {{\theta _k}} \right) + {\bm{b}}\left( {{\theta _k}} \right){{\bm{a}}^{\rm{T}}}\left( {{\theta _0}} \right)} \right)} \right){\bm{s}} \vspace{2pt}\\
\quad \   = {{{{\bm {A}}}}_k}{\bm{s}}
\end{array}
\label{MPTargetReceiveEqReciprocity}
\end{equation}

To this end, it is worth pointing out that the scattering reciprocities of the target and scatterers only impact on the dimension of $\bm{u}$, instead of the solving algorithms. Without loss of generality, we assume that the scattering reciprocities hold in our following discussion, i.e., (\ref{MPTargetReceiveEqReciprocity}) is used to model ${{{\bm{y}}}_{k}}$.

\section{Algorithms for the spherical uncertainty set}
In this section, we devise the algorithms for robust waveform design problem against the spherical uncertainty set, namely ${{\cal P}_1}$ described in (\ref{RobustSINRType1}). We prove that the optimal filter can be analytically calculated first. Then, the inner minimization problem with respect to $\bm{u}$ is converted to a maximization problem based on Lagrange duality\cite{BoydConvex,NemirovskiLectures}. Thus, the max-min ${{\cal P}_1}$ problem can be reformulated as a maximization problem with respect to $\bm{s}$ and the dual variable of $\bm{u}$. SDR method is used\cite{LuoSemidefinite} to approximate the non-convex maximization problem with a Semi-Definite Programming(SDP) problem. Therefore, we call this algorithm Duality Maximization Semi-Definite Relaxation(DMSDR). Then, the synthesis schemes of transmit waveform and receive filter pair ($\bm{s}^*, \bm{w}^*$) are provided. Finally, we analysize the convergences and computational complexities of the proposed MDSDR algorithm briefly.
\subsection{Problem reformulation of ${{\cal P}_1}$}
The following proposition provide the basic properties of ${{\cal P}_1}$.
\newtheorem{propositions}{Proposition}
\begin{propositions}
The optimal solution of ${{\cal P}_1}$ is equivalent to the following optimization problem.
\begin{equation}
{{\cal P}'_1}\left\{ \begin{array}{l}
\mathop {{\rm{max}}}\limits_{{\bm{s}}} \mathop {{\rm{ min}}}\limits_{\bm{u}} \mathop {{\rm{max}}}\limits_{{\bm{w}}}{\rm{ }}\frac{{{{\left| {{{\bm{w}}^{\rm{H}}}{\bm{Yu}}} \right|}^2}}}{{{{\bm{w}}^{\rm{H}}}{{\bm{R}}_n}{\bm{w}}}}\\
s.t.\quad {\left\| {{\bm{u}} - {{\bm{u}}_0}} \right\|_2} \le r,\left| {{\bm{s}(i)}} \right| = 1
\end{array} \right.
\end{equation} 
\end{propositions}
\begin{IEEEproof}
	See Appendix A.
\end{IEEEproof}

Note that the optimal $\bm{w}^*$ for any $\bm{s}$ and $\bm{u}$ can be represented as $\bm{w}^*={\bm{R}}_n^{ - 1}{\bm{Yu}}$. As an immediate consequence of \textit{Proposition 1}, we reformulate ${{\cal P}_1}$ by solving the optimal filter $\bm{w}$ and rewrite it as 
\begin{equation}
{{\cal P}_1}\left\{ \begin{array}{l}
\mathop {{\rm{max}}}\limits_{\bm{s}} \mathop {{\rm{ min}}}\limits_{\bm{u}} {\rm{ }}{{\bm{u}}^{\rm{H}}}{{\bm{Y}}^{\rm{H}}}{\bm{R}}_n^{ - 1}{\bm{Yu}}\\
s.t.\quad {\left\| {{\bm{u}} - {{\bm{u}}_0}} \right\|_2} \le r,\\
\quad \ \ \ \left| {{\bm{s}(i)}} \right| = 1, i=1,2,..LN_T
\end{array} \right..
\label{MPTargetReceiveEqFirstTypeSolvingW}
\end{equation}
Next, we consider the inner minimization problem
\begin{equation}
{{\cal P}_{1,1}}\left\{ \begin{array}{l}
\mathop {{\rm{ min}}}\limits_{\bm{u}} {\rm{ }}{{\bm{u}}^{\rm{H}}}{{\bm{Y}}^{\rm{H}}}{\bm{R}}_n^{ - 1}{\bm{Yu}}\\
s.t.\quad {\left\| {{\bm{u}} - {{\bm{u}}_0}} \right\|_2^2} \le r^2
\end{array} \right..
\label{InnerMinimizationFirstType}
\end{equation}
It is easy to find that the optimization problem described in (\ref{InnerMinimizationFirstType}) is a convex problem and meets the Slater condition\cite{BoydConvex}. Thus, the strong duality holds and we can express  (\ref{InnerMinimizationFirstType}) as a dual form based on the following proposition.

\begin{propositions}
The dual problem of ${{\cal P}_{1,1}}$ is
\begin{equation}
\begin{array}{l}
\mathop {{\rm{max}}}\limits_\mu  {\rm{ }} - {\mu ^2}{\bm{u}}_0^{\rm{H}}{\left( {{{\bm{Y}}^{\rm{H}}}{\bm{R}}_n^{ - 1}{\bm{Y}} + \mu \bf{I}} \right)^{ - 1}}{\bm{u}}_0 + \mu \left( {{\bm{u}}_0^{\rm{H}}{{\bm{u}}_0} - {r^2}} \right)\\
s.t.\quad \mu  \ge 0
\end{array},
\end{equation}
where $\mu$ is the corresponding dual variable. 
\end{propositions}
\begin{IEEEproof}
	See Appendix B.
\end{IEEEproof}

According to \textit{Proposition 2}, we can easily reformulate the max-min problem in (\ref{MPTargetReceiveEqFirstTypeSolvingW}) as an equivalent maximization problem with respect to $\bm{s}$ and $\mu$,
\begin{equation}
{{\cal P}_2}\left\{ \begin{array}{l}
\mathop {{\rm{max}}}\limits_{{\bm{s}},\mu } {\rm{ }} - {\mu ^2}{\bm{u}}_0^{\rm{H}}{\left( {{{\bm{Y}}^{\rm{H}}}{\bm{R}}_n^{ - 1}{\bm{Y}} + \mu \bf{I}} \right)^{ - 1}}{\bm{u}}_0^{} + \mu \left( {{\bm{u}}_0^{\rm{H}}{{\bm{u}}_0} - {r^2}} \right)\\
s.t.\quad \mu  \ge 0,\left| {{\bm{s}(i)}} \right| = 1, i=1,2,...LN_T.
\end{array} \right.
\label{FirstTypeMaxMax}
\end{equation}

\subsection{Optimization algorithms for ${{\cal P}_2}$}
Let ${\bm{T}}({\bm{s}}) = {{\bm{Y}}^{\rm{H}}}{\bm{R}}_n^{ - 1}{\bm{Y}}$, then ${{\cal P}_2}$ is equivalent to the following problem by introducing an auxiliary variable $t$ and converting the maximization problem into a minimization problem,
\begin{equation}
{{\cal P}_3}\left\{ \begin{array}{l}
\mathop {{\rm{min }} \  t}\limits_{{\bm{s}},\mu ,t} \\
s.t.\quad \left[ {\begin{array}{*{20}{c}}
	{t + \mu \left( {{\bm{u}}_0^{\rm{H}}{{\bm{u}}_0} - {r^2}} \right)}&{\mu {\bm{u}}_0^{\rm{H}}}\\
	{\mu {\bm{u}}_0^{}}&{{\bm{T}}({\bm{s}}) + \mu \bf{I}}
	\end{array}} \right] \succeq 0\\
\quad \quad \mu  \ge 0,\left| {\bm{s}(i)} \right| = 1, i=1,2,...LN_T
\end{array} \right..
\label{FirstTypeMinSchur}
\end{equation}  
\begin{propositions}
Define the covariance matrix  ${\bm{R}_s} = {\bm{s}}{\bm{s}^{\rm{H}}}$, then ${\bm{T}({\bm{s})}}$ is a linear function with respect to ${\bm{R}_s}$ with the ($i$,$j$)th element
\begin{equation}
{\bm{T}}\left( {i,j} \right) = {\rm{tr}}\left( {{\bm{A}}_{i - 1}^{\rm{H}}{\bm{R}}_n^{ - 1}{{\bm{A}}}_{j - 1}{{\bm{R}}_s}} \right).
\end{equation}
\end{propositions} 
\begin{IEEEproof}
	See Appendix C.
\end{IEEEproof}

However, ${\cal{P}}_3$ is still NP hard due to the CMC constraint on $\bm{s}$\cite{SoltanalianDesigning}. Inspired by \textit{Prposition 3}, we optimize ${{\cal P}_3}$ with respect to $\left( {\bm{R}_s}, \mu, t\right) $, and relax it by dropping the rank constraint on ${\bm{R}_s}$. To this end, the SDR form of ${{\cal P}_3}$ is given by 
\begin{equation}
{{\cal P}_4}\left\{ \begin{array}{l}
\mathop {{\rm{min }} \ t}\limits_{{{\bm{R}}_s},\mu ,t} \\
s.t.\quad \left[ {\begin{array}{*{20}{c}}
	{t + \mu \left( {{\bm{u}}_0^{\rm{H}}{{\bm{u}}_0} - {r^2}} \right)}&{\mu {\bm{u}}_0^{\rm{H}}}\\
	{\mu {\bm{u}}_0^{}}&{{\bm{T}} + \mu \bf{I}}
	\end{array}} \right] \succeq 0\\
\quad \quad \mu  \ge 0,{\rm{diag}}({{\bm{R}}_s}) = {\bf{1}}, {\bm{R}_s} \succeq 0
\end{array} \right..
\label{OptimizationRsSpherical}
\end{equation}
It is easy to verify the convexity of ${{\cal P}_4}$ due to the linear objective as well as linear constraints. Thus, it can be solved in polynomial time by CVX toolbox\cite{GrantCVX}.
%
%
\subsection{Synthesize transmit waveform and receive filter from $\bm{R}_s^*$}
Denoted $\left( \bm{R}_s^*,\mu^*,t^*\right) $ by the optimal solution of ${\cal{P}}_4$. Let us consider the synthesis of transmit waveform and receive filter pair ($\bm{s}^*$,$\bm{w}^*$) from \vspace{3pt} $\bm{R}_s^*$. If $\bm{R}_s^*$ is rank-one, we can directly synthesize $\bm{s}^*$ by $\bm{R}_s^*=\bm{s}^*\left( \bm{s}^*\right) ^{\rm{H}}$. As to $\bm{w}^*$, we solve the following optimization problem
\begin{equation}
{\bm{u}^*} = \mathop {{\rm{argmin}}}\limits_{\left\{ {{\bm{u}}\left| {{{\left\| {{\bm{u}} - {{\bm{u}}_0}} \right\|}_2} \le r} \right.} \right\}} {\rm{ }}{{\bm{u}}^{\rm{H}}}{\bm{T}}({{\bm{s}}^*}){\bm{u}},
\label{SynthesizeURankOne}
\end{equation} 
and synthesize it with
\begin{equation}
{{\bm{w}}^*} = {\bm{R}}_n^{ - 1}{\bm{Y}}({{\bm{s}}^*}){\bm{u}}^*.
\label{SynthesizeFilterRankOne}
\end{equation}

Otherwise we leverage on the randomization schemes\cite{LuoSemidefinite} to generate the transmit waveform. In particular, we draw $Q$ random vectors $\bm{v}_1,\bm{v}_2,...,\bm{v}_Q$ from the complex Gaussian distribution ${\cal CN}({\bf{0}},{{\bm{R}}_s^*})$, and synthesize $\bm{s}^{(q)}$ by ${{\bm{s}}^{(q)}}={e^{j\arg ({{\bm{v}}_q})}}$. Then, we compute the minimum output SINR ${\gamma _q}$ with
\begin{equation}
{\gamma _q} = \mathop {{\rm{min}}}\limits_{\left\{ {{\bm{u}}\left| {{{\left\| {{\bm{u}} - {{\bm{u}}_0}} \right\|}_2} \le r} \right.} \right\}} {\rm{ }}{{\bm{u}}^{\rm{H}}}{\bm{T}}({{\bm{s}}^{(q)}}){\bm{u}},
\label{SynthesizeU}
\end{equation}
and record the corresponding optimal solution $\bm{u}_q^*$. Pick the maximum value in $\left\lbrace {\gamma _1},{\gamma _2},...,{\gamma _Q}\right\rbrace $, for example $\gamma_q$, then we synthesize transmit waveform and receive filter pair ($\bm{s}^*$,$\bm{w}^*$) with
\begin{equation}
{{\bm{s}}^*} = {{\bm{s}}^{(q)}},{{\bm{w}}^*} = {\bm{R}}_n^{ - 1}{\bm{Y}}({{\bm{s}}^*}){\bm{u}}_i^*.
\label{SynthesizeFilter}
\end{equation}

In order to give a clear expression, Table \ref{AlgorithmDMSDR} summarizes the DMSDR algorithm. 
\begin{table}[!t]
	\renewcommand{\arraystretch}{1.3}
	\caption{DMSDR for the sperical uncertainty set}
	\label{AlgorithmDMSDR}
	\centering
	\begin{tabular}{l}
		\hline
		\textbf{Input}:
		\@ \@ $\left\{ {{{\bm{A}}_k}} \right\}_{k = 0}^K$,
		 ${{\bm{R}}_{n}}$, ${{\bm{u}}_{0}}$ and $r$.\\ 
		\hline
		\@ \@\@ \@ \textit{Step 1}: Get ${\bm{R}}_s^*$ by  solving ${\cal{P}}_4$;\\
		\@ \@\@ \@ \textit{Step 2}: Synthesize ${\bm{s}^{*}}$ and ${\bm{w}^{*}}$ from ${\bm{{R}}_s^{*}}$. If ${\rm{rank}({\bm{{R}}_s^{*}})} =1$,\\
		\@ \@\@ \@\@ \@\@ \@\@ \@\@ \@ \@ \@\@ \@ \@ \@ ${\bm{R}}_s^{(*)} = {{\bm{s}}^{*}}{\left( {{{\bm{s}}^{*}}} \right)^{\rm{H}}}$, then synthesize ${\bm{w}^{*}}$ with (\ref{SynthesizeURankOne}) and (\ref{SynthesizeFilterRankOne});\\
		\@ \@\@ \@\@ \@\@ \@\@ \@\@ \@ \@ \@\@ \@ \@ \@ otherwise the randomization schemes are used to generate  ${\bm{s}^{*}}$\\
		\@ \@\@ \@\@ \@\@ \@\@ \@\@ \@ \@ \@\@ \@ \@ \@ and ${\bm{w}^{*}}$ with (\ref{SynthesizeU}) and (\ref{SynthesizeFilter});\\
		\textbf{Output}: 
		\@ \@ ${\bm{s}^{*}}$ and ${\bm{w}^{*}}$.\\
		\hline
	\end{tabular}
\end{table}
\subsection{Further discussions on DMSDR}
In this subsection, we give some discussions on convergences and computational complexities of DMSDR. 

Obviously, it is easy to verify the convergences of DMSDR due to the convexity of ${{\cal P}_4}$, and the locally optimal solution
also means the globally optimal solution.

As to the computational complexities, it requires at most $O\left(\left( LN_T\right)^{6.5} + \left( LN_T\right)^{4}K^{2.5} + 
\left( LN_T\right)^{2}K^{3.5}\right)$ operations to solve ${{\cal P}_4}$\cite{NemirovskiLectures}.
And $O\left(Q\left( LN_T\right)^{2} + QK^{3}\right)$ operations are needed to generate $\bm{s}^{(i)}$ with randomization\cite{aubry2012cognitive}, i.e., it requires\vspace{2pt} $O\left(\left( LN_T\right)^{2}\right)$ operations to generate $\bm{s}^{(i)}$ and $O\left(K^{3}\right)$ operations to solve  (\ref{SynthesizeU})\cite{li2003robust}. Additionally,\vspace{2pt}, $O\left(K\left( LN_R\right)^{2} \right)$ operations are needed to compute $\bm{w}^{*}$ with (\ref{SynthesizeFilter}). Note that, in most practical\vspace{2pt} situations, the number $\left( LN_T\right)^{6.5}$ takes the dominance, \vspace{2pt}thus the total \vspace{2pt} computational complexities of DMSDR are given by $O\left(\left( LN_T\right)^{6.5}\right)$.

\section{Algorithms for the annular uncertainty set}
This section is devoted to the algorithms for robust waveform design problem against the annular uncertainty set, namely ${\tilde {\cal P}_1}$ described in (\ref{RobustSINRType2}). Similarly to the former situation, the inner minimization problem with respect to $\bm{u}$ is considered firstly. Unfortunately, the inner minimization is difficult to deal with due to the non-convex constraint sets. We use the SDR method by letting ${{\bm{R}_u}=\bm{u}\bm{u}^{\rm{H}}} $ and dropping the rank constraint to approximate the inner minimization problem with a SDP problem. Thus, the max-min problem can be expressed as a maximization problem based on Lagrange duality. Further, the SDR method is used again to get the optimal transmit and receive covariance matrix, and we call the algorithm Duality Maximization Double Semi-Definite Relaxation(DMDSDR). Then, the synthesis schemes of transmit waveform and receive filter pair ($\bm{s}^*, \bm{w}^*$) are provided. Finally, the convergences and computational complexities of DMDSDR are analysized.   
\subsection{Problem reformulation of ${\tilde{{\cal P}}_1}$}

Now, we consider the inner minimization problem
\begin{equation}
{\tilde {\cal P}_{1,1}}\left\{ \begin{array}{l}
\mathop {{\rm{min}}}\limits_{\bm{u}} {\rm{ }}{{\bm{w}}^{\rm{H}}}{\bm{Yu}}{{\bm{u}}^{\rm{H}}}{{\bm{Y}}^{\rm{H}}}{\bm{w}}\\
s.t.\quad  {\bm{\eta}(k)} \le \left| {{\bm{u}(k)}} \right| \le {\bm{\xi}(k)},k = 1,2,...,K + 1
\end{array} \right..
\label{InnerMinimizationSecondType}
\end{equation}

The problem described in (\ref{InnerMinimizationSecondType}) is also NP hard due to the modulus constraints on $\bm{u}$. Let ${{\bm{R}_u}=\bm{u}\bm{u}^{\rm{H}}} $, we study its SDP form  by dropping the rank constraint on ${\bm{R}_u}$, namely,
\begin{equation}
{\tilde {\cal P}_{1,2}}\left\{ \begin{array}{l}
\mathop {{\rm{min}}}\limits_{{{\bm{R}}_u}} {\rm{ tr}}\left( {{{\bm{Y}}^{\rm{H}}}{\bm{w}}{\bm{w}}^{\rm{H}}{\bm{Y}}{{\bm{R}}_u}} \right)\\
s.t.\quad { {{\bm{\eta}}(k)^2} } \le {{\bm{R}}_u}(k,k) \le { {{\bm{\xi}(k)^2}}},k=1,2,...K+1\\
\quad \quad{\bm{R}_u} \succeq 0
\end{array} \right..
\label{InnerMinimizationSecondTypeSDR}
\end{equation}	 
${\tilde {\cal P}_{1,2}}$ is a SDP problem with linear objective as well as linear constraints, and one can easy find the its dual problem based on the following proposition.

\begin{propositions}
 The dual problem of ${\tilde {\cal P}_{1,2}}$ is
\begin{equation}
{\tilde{\cal{P}}_{1,3}}\left\{ \begin{array}{l}
\mathop {{\rm{max}}}\limits_{{\bm{\mu }},{\bm{h}},{\bm{Z}}} {\rm{ }}{{\bm{u}}^{\rm{T}}}{\bm{f} - }{{\bm{h}}^{\rm{T}}}{\bm{g}}
\\
s.t.\ {{\bm{Y}}^{\rm{H}}}{\bm{w}}{\bm{w}^{\rm{H}}}{\bm{Y}} - {\bm{Z}} + \sum\limits_{k = 1}^{K + 1} {\left( { {\bm{h}(k)}-{\bm{\mu}(k)}} \right){{\bm{E}}_k}}  = {\bf{0}}\\
\quad \quad {{\bm{\mu}(k)}} \ge 0,{\bm{h}(k)} \ge 0, k=1,2,3...,K+1\\
\quad \quad {\bm{Z}} \succeq 0
\end{array} \right.,
\end{equation}
where $\bm{\mu}, \bm{h}, \bm{Z}$ are the corresponding dual variables. And ${\bm{f}} = {\bm{\eta }} \odot {\bm{\eta }}$, ${\bm{g}} = {\bm{\xi }} \odot {\bm{\xi }}$
, $\bm{E}_k = {\bm{e}_k}{\bm{e}_k^{\rm{H}}}$.
\end{propositions}
\begin{IEEEproof}
	See Appendix D.
\end{IEEEproof}

Based on \textit{Proposition 4}, we reformulate the max-min problem ${\tilde {\cal P}_1}$ by converting the inner minimization problem to its dual problem with SDR. Then we get 
\begin{equation}
{\tilde {\cal P}_2}\left\{ \begin{array}{l}
\mathop {{\rm{max}}}\limits_{{\bm{\mu }},{\bm{h}},{\bm{Z}},{\bm{s}},{\bm{w}}} {\rm{ }}\frac{{{{\bm{u}}^{\rm{T}}}{\bm{f} - }{{\bm{h}}^{\rm{T}}}{\bm{g}}}}{{{{\bm{w}}^{\rm{H}}}{{\bm{R}}_n}{\bm{w}}}}{\rm{ }}\\
s.t.\  {{\bm{Y}}^{\rm{H}}}{\bm{w}}{{\bm{w}}^{\rm{H}}}{\bm{Y}} - {\bm{Z}} + \sum\limits_{k = 1}^{K + 1} {\left( {{\bm{h}(k)} - {\bm{\mu}(k)}} \right){{\bm{E}}_k}}  = {\bf{0}}\\
\quad \ {\bm{\mu}(k)} \ge 0,{\bm{v}(k)} \ge 0,k = 1,2,...,K + 1\\
\quad \ {\bm{Z}} \succeq 0\\
\quad \ \left| {{\bm{s}(i)}} \right| = 1,i = 1,2,...,L{N_T}
\end{array} \right..
\label{MaxDualitySecondType}
\end{equation}

Next, let us investigate the relationships between  ${\tilde{\cal{P}}_2}$ and  ${\tilde{\cal{P}}_1}$ briefly. For any given $\bm{w}$ and $\bm{s}$, the following two equalities(inequalities) hold
\begin{equation}
\mathop {{\rm{max}}}\limits_{{\bm{\mu }},{\bm{h}},{\bm{Z}}} {\rm{ }}\frac{{{{\bm{u}}^{\rm{T}}}{\bm{f} - }{{\bm{h}}^{\rm{T}}}{\bm{g}} }}{{{{\bm{w}}^{\rm{H}}}{{\bm{R}}_n}{\bm{w}}}} = \mathop {\min }\limits_{{{\bm{R}}_u}} {\rm{ }}\frac{{{\rm{tr}}\left( {{{\bm{Y}}^{\rm{H}}}{\bm{w}}{{\bm{w}}^{\rm{H}}}{\bm{Y}}{{\bm{R}}_u}} \right)}}{{{{\bm{w}}^{\rm{H}}}{{\bm{R}}_n}{\bm{w}}}},
\label{StrongDualitEquality}
\end{equation}
\begin{equation}
\mathop {\min }\limits_{{{\bm{R}}_u}} {\rm{ }}\frac{{{\rm{tr}}\left( {{{\bm{Y}}^{\rm{H}}}{\bm{w}}{{\bm{w}}^{\rm{H}}}{\bm{Y}}{{\bm{R}}_u}} \right)}}{{{{\bm{w}}^{\rm{H}}}{{\bm{R}}_n}{\bm{w}}}} \le \mathop {\min }\limits_{\bm{u}} \frac{{{{\bm{u}}^{\rm{H}}}{{\bm{Y}}^{\rm{H}}}{\bm{w}}{{\bm{w}}^{\rm{H}}}{\bm{Yu}}}}{{{{\bm{w}}^{\rm{H}}}{{\bm{R}}_n}{\bm{w}}}},
\label{RelaxtionInequality}
\end{equation}
where $\bm{\mu},\bm{h},\bm{Z}$ belong to the feasible set of ${\tilde{\cal{P}}_{1,3}}$, $\bm{R}_u$ belongs to the feasible set of ${\tilde{\cal{P}}_{1,2}}$ and $\bm{u}$ belongs to the feasible set of ${\tilde{\cal{P}}_{1,1}}$. (\ref{StrongDualitEquality}) holds due to the strong duality between ${\tilde{\cal{P}}_{1,2}}$ and ${\tilde{\cal{P}}_{1,3}}$, while the reason for (\ref{RelaxtionInequality}) is that the feasible set of ${\tilde{\cal{P}}_{1,1}}$ is included by the feasible set of ${\tilde{\cal{P}}_{1,2}}$.

Combining  (\ref{StrongDualitEquality}) and (\ref{RelaxtionInequality}), we know that for any given transmit waveform and receive filter pair $(\bm{s},\bm{w})$, ${\tilde{\cal{P}}_{1,3}}$ provides a lower bound of the worst-case SINR with respect to $\bm{u}$. Thus, we can explain ${\tilde{\cal{P}}_{2}}$ that we maximize the lower bound of worst-case SINR by designing $(\bm{s},\bm{w})$.   

However, ${\tilde{\cal{P}}_{2}}$ is also difficult to solve directly due to the quadratic equality constraints on $\bm{w}$ and $\bm{s}$. In order to solve ${\tilde{\cal{P}}_{2}}$ efficiently, we adopt the SDR method again with $\bm{W}={\bm{w}}{\bm{w}}^{\rm{H}}$ and $\bm{R}_s={\bm{s}}{\bm{s}}^{\rm{H}}$. To this end, we reformulate ${\tilde{\cal{P}}_{2}}$ as
\begin{equation}
{\tilde {\cal P}_3}\left\{ \begin{array}{l}
\mathop {{\rm{max}}}\limits_{{\bm{\mu }},{\bm{h}},{\bm{Z}},{{\bm{R}}_s},{\bm{W}}} {\rm{ }}\frac{{{{\bm{u}}^{\rm{T}}}{\bm{f} - }{{\bm{h}}^{\rm{T}}}{\bm{g}} }}{{{\rm{tr}}\left( {{{\bm{R}}_n}{\bm{W}}} \right)}}{\rm{ }}\\
s.t.\quad {{\bm{Y}}^{\rm{H}}}{\bm{WY}} - {\bm{Z}} + \sum\limits_{k = 1}^{K + 1} {\left( {{\bm{h}(k)} - {\bm{\mu}(k)}} \right){{\bm{E}}_k}}  = {\bf{0}}\\
\quad \quad{\bm{\mu}(k)} \ge 0,{\bm{h}(k)} \ge 0,k = 1,2,...,K + 1\\
\quad \quad{\bm{Z}} \succeq 0,{\bm{W}} \succeq 0,{{\bm{R}}_s} \succeq 0\\
\quad \quad{\rm{diag}}({{\bm{R}}_s}) = {\bf{1}}
\end{array} \right..
\label{MaxSDRSecondType}
\end{equation}

We resort to a cyclic optimization method to solve ${\tilde {\cal P}_3}$. More exactly, we initialize the transmit waveform covariance $\bm{R}_s^{(0)}$, then alternatively maximize the objective with respect to $(\bm{\mu},\bm{h},\bm{Z}, \bm{W}^{(m-1)})$ for a fixed $\bm{R}_s^{(m-1)}$ and maximize the objective with respect to $(\bm{\mu},\bm{h},\bm{Z}, \bm{R}_s^{(m)})$ for a fixed ${\bm{W}^{(m-1)}}$.

\subsection{Optimization for ${\bm{W}^{(m)}}$}
For a fixed $\bm{R}_s^{(m)}$, the optimization problem for ${\bm{W}^{(m)}}$ is given by
\begin{equation}
{\tilde {\cal P}_{\bm{W}}^{(m)}}\left\{ \begin{array}{l}
\mathop {{\rm{max}}}\limits_{{\bm{\mu }},{\bm{h}},{\bm{Z}},{\bm{W}}} {\rm{ }}\frac{{{\bm{u}}^{\rm{T}}}{\bm{f} - }{{\bm{h}}^{\rm{T}}}{\bm{g}}}{{{\rm{tr}}\left( {{{\bm{R}}_n}{\bm{W}}} \right)}}{\rm{ }}\\
s.t.\quad {\bm{G(W)}} - {\bm{Z}} + \sum\limits_{k = 1}^{K + 1} {\left( {{\bm{h}(k)} - {\bm{\mu}(k)}} \right){{\bm{E}}_k}}  = {\bf{0}}\\
\quad \quad {\bm{\mu}(k)} \ge 0,{\bm{h}(k)} \ge 0,k = 1,2,...,K + 1\\
\quad \quad {\bm{Z}} \succeq 0,{\bm{W}} \succeq 0
\end{array} \right.,
\label{OptimizaWSecondType}
\end{equation} 
where $\bm{G(W)}={\bm{Y}}^{\rm{H}}{\bm{WY}}$ is a linear function with respect to $\bm{W}$
with ${\bm{G}}\left( {i,j} \right) = {\rm{tr}}\left( {{\bm{A}}_{i - 1}^{\rm{H}}{\bm{W}}{{\bm{A}}}_{j - 1}{{\bm{R}}_s^{(m)}}} \right)$(see \textit{Proposition 3}). Define the following problem, 
\begin{equation}
{\tilde {\cal P}_{\bm{W}}^{(m)\prime}}\left\{ \begin{array}{l}
\mathop {{\rm{max}}}\limits_{{\bm{\mu }},{\bm{h}},{\bm{Z}},{\bm{W}}} {\rm{ }}{{\bm{u}}^{\rm{T}}}{\bm{f} - }{{\bm{h}}^{\rm{T}}}{\bm{g}}\\
s.t.\quad {\bm{G(W)}} - {\bm{Z}} + \sum\limits_{k = 1}^{K + 1} {\left( {{\bm{h}(k)} - {\bm{\mu}(k)}} \right){{\bm{E}}_k}}  = {\bf{0}}\\
\quad \quad{\bm{\mu}(k)} \ge 0,{\bm{h}(k)} \ge 0,k = 1,2,...,K + 1\\
\quad \quad{\bm{Z}} \succeq 0,{\bm{W}} \succeq 0\\
\quad \quad{\rm{tr}}\left( {{{\bm{R}}_n}{\bm{W}}} \right) = 1
\end{array} \right.,
\label{OptimumWSecondType}
\end{equation}
and \textit{Proposition 5} shows the relationships between ${\tilde {\cal P}_{\bm{W}}^{(m)}}$ and ${\tilde {\cal P}_{\bm{W}}^{(m)\prime}}$.

\begin{propositions}
Let the optimal value for ${\tilde {\cal P}_{\bm{W}}^{(m)\prime}}$ be $p_{\bm{W}}^{(m)\prime}$ with optimal solution $(\bm{\mu}^*,\bm{h}^*,\bm{Z}^*,\bm{W}^{(m)})$, and the optimal value for ${\tilde {\cal P}_{\bm{W}}^{(m)}}$ be $p_{\bm{W}}^{(m)}$, then $p_{\bm{W}}^{(m)\prime}=p_{\bm{W}}^{(m)}$  \vspace{1pt} and $(\bm{\mu}^*,\bm{h}^*,\bm{Z}^*,\bm{W}^{(m)})$ is also the optiaml solution for ${\tilde {\cal P}_{\bm{W}}^{(m)}}$.
\end{propositions}
\begin{IEEEproof}
	See Appendix E.
\end{IEEEproof}

According to \textit{Proposition 5}, for a given $\bm{R}_s^{(m)}$, we optimize ${\tilde {\cal P}_{\bm{W}}^{(m)\prime}}$ to get $\bm{W}^{(m)}$, which is a convex optimization problem and can be solved in polynomial time.   

\subsection{Optimization for ${\bm{R}_s^{(m)}}$}
For a fixed ${\bm{W}^{(m-1)}}$, the optimization problem for ${\bm{R}_s^{(m)}}$ is given by
\begin{equation}
{\tilde {\cal P}_{{{\bm{R}}_s}}^{(m)}}\left\{ \begin{array}{l}
\mathop {{\rm{max}}}\limits_{{\bm{\mu }},{\bm{h}},{\bm{Z}},{{\bm{R}}_s}} {\rm{ }}\frac{{{{\bm{u}}^{\rm{T}}}{\bm{f} - }{{\bm{h}}^{\rm{T}}}{\bm{g}} }}{{{\rm{tr}}\left( {{{\bm{R}}_n}{\bm{W}^{(m-1)}}} \right)}}{\rm{ }}\\
s.t.\quad {\bm{\tilde G}}({{\bm{R}}_s}) - {\bm{Z}} + \sum\limits_{k = 1}^{K + 1} {\left( {{\bm{h}(k)} - {\bm{\mu}(k)}} \right){{\bm{E}}_k}}  = {\bf{0}}\\
{\bm{\mu}(k)} \ge 0,{\bm{h}(k)} \ge 0,k = 1,2,...,K + 1\\
{\bm{Z}} \succeq 0,{{\bm{R}}_s} \succeq 0\\
{\rm{diag}}({{\bm{R}}_s}) = {\bf{1}}
\end{array} \right.,
\label{OptimizeRsSecondType}
\end{equation}
where ${\bm{\tilde G}}({{\bm{R}}_s})$ is a linear function with respect to $\bm{R}_s$
with ${\bm{\tilde{G}}}\left( {i,j} \right) = {\rm{tr}}\left( {{\bm{A}}_{i - 1}^{\rm{H}}{\bm{W}^{(m-1)}}{ {\bm{A}}}_{j - 1}{{\bm{R}}_s}} \right)$(see \textit{Proposition 3}).
It is easy to verify the convexity of ${\tilde {\cal P}_{{{\bm{R}}_s}}^{(m)}}$, so it can be efficiently solved in polynomial time too. 

\subsection{Synthesize transmit waveform and receive filter from ($\bm{R}_s^*$,$\bm{W}^*$)}
The remaining problem is to synthesize transmit waveform and receive filter pair ($\bm{s}^*$,$\bm{w}^*$) from ($\bm{R}_s^*$,$\bm{W}^*$). If $\bm{R}_s^*$ or $\bm{W}^*$ is rank-one, we can directly \vspace{2pt} synthesize ($\bm{s}^*$,$\bm{w}^*$) by $\bm{R}_s^*=\bm{s}^*\left( \bm{s}^*\right) ^{\rm{H}}$ or $\bm{W}^*=\bm{w}^*\left( \bm{w}^*\right) ^{\rm{H}}$. Otherwise, we use randomization method to approximate $\bm{R}_s^*$ and $\bm{W}^*$.

Similarly to the DMSDR algorithm, we draw $Q$ random vectors $\bm{v}_1,\bm{v}_2,...,\bm{v}_Q$ from the complex Gaussian distribution ${\cal CN}({\bf{0}},{{\bm{R}}_s^*})$, and synthesize $\bm{s}^{(q)}$ with ${{\bm{s}}^{(q)}}={e^{j\arg ({{\bm{v}}_{q}})}}$. Then we compute the minimum output SINR ${\gamma _q}$ with

\begin{equation}
\begin{array}{l}
{\gamma _q} = \mathop {{\rm{min}}}\limits_{{{\bm{R}}_u}} {\rm{ }}\frac{{{\rm{tr}}\left( {{\bm{Y}}{{({{\bm{s}}^{(q)}})}^{\rm{H}}}{{\bm{W}}^*}{\bm{Y}}({{\bm{s}}^{(q)}}){{\bm{R}}_u}} \right)}}{{{\rm{tr}}\left( {{{\bm{W}}^*}{{\bm{R}}_n}} \right)}}\\
s.t.\quad {\bm{f}(k)} - {\rm{tr}}\left( {{{\bm{E}}_k}{{\bm{R}}_u}} \right) \le 0,\\
\quad \quad \;{\rm{tr}}\left( {{{\bm{E}}_k}{{\bm{R}}_u}} \right) - {\bm{g}(k)} \le 0, k=1,2,...,K+1\\
\quad \quad \;{{\bm{R}}_u} \succeq 0
\end{array}.
\label{SynthesizeSworstSINR}
\end{equation}
Pick the maximum value in $\left\lbrace {\gamma _1},{\gamma _2},...,{\gamma _Q}\right\rbrace $, for example $\gamma_q$, then we synthesize transmit waveform with 
\begin{equation}
{{\bm{s}}^*} = {{\bm{s}}^{(q)}}.
\label{SynthesizeS}
\end{equation}
As to $\bm{w}^*$, we also draw $Q$ random vectors $\bm{w}_1,\bm{w}_2,...,\bm{w}_Q$ 
from the complex Gaussian distribution ${\cal CN}({\bf{0}},{{\bm{W}}^*})$. Then we compute the minimal output SINR ${\tilde{\gamma} _{q}}$ with	
\begin{equation}
\begin{array}{l}
{\tilde{\gamma _{q}}} = \mathop {{\rm{min}}}\limits_{{{\bm{R}}_u}} {\rm{ }}\frac{{{\rm{tr}}\left( {{\bm{Y}}{{({{\bm{s}}^*})}^{\rm{H}}}{{\bm{w}_q}{\bm{w}_q^{\rm{H}}}}{\bm{Y}}({{\bm{s}}^*}){{\bm{R}}_u}} \right)}}{{{\rm{tr}}\left( {{{\bm{w}_q}{\bm{w}_q^{\rm{H}}}}{{\bm{R}}_n}} \right)}}\\
s.t.\quad {\bm{f}(k)} - {\rm{tr}}\left( {{{\bm{E}}_k}{{\bm{R}}_u}} \right) \le 0,\\
\quad \quad \;{\rm{tr}}\left( {{{\bm{E}}_k}{{\bm{R}}_u}} \right) - {\bm{g}(k)} \le 0, k=1,2,...,K+1\\
\quad \quad \;{{\bm{R}}_u} \succeq 0
\end{array}.
\label{SynthesizeWworstSINR}
\end{equation}	
Pick the maximum value in $\left\lbrace {\tilde{\gamma} _1},{\tilde{\gamma} _2},...,{\tilde{\gamma} _Q}\right\rbrace $, for example $\tilde{\gamma}_q$, then we synthesize receive filter with 
\begin{equation}
{{\bm{w}}^*} = {{\bm{w}}_q} .
\label{SynthesizeW}
\end{equation}	
	
In order to give a clear expression, Table \ref{AlgorithmDMDSDR} summarizes the DMDSDR algorithm. 
\begin{table}[!t]
	\renewcommand{\arraystretch}{1.3}
	\caption{DMDSDR for the annular uncertainty set}
	\label{AlgorithmDMDSDR}
	\centering
	\begin{tabular}{l}
		\hline
		\textbf{Input}:
		\@ \@ $\left\{ {{{\bm{A}}_k}} \right\}_{k = 0}^K$,
		${{\bm{R}}_{n}}$, $\bm{\eta}$, $\bm{\xi}$ and $\varepsilon$.\\ 
		\hline
		\textbf{Initialization}: \\
		\@ \@\@ \@ Set $m=0$, initialize the transmit signal ${\bm{s}}^{(0)}$. \\
		\textbf{Iteration}:\\
		\@ \@\@ \@ \textit{Step 1}: Optimize $\bm{W}^{(m)}$ with (\ref{OptimumWSecondType}), and record the optimal value\\
		\@ \@\@ \@ \@ \@\@ \@ \@ \@\@ \@ \@ \@\@ \@ as $\tilde{\gamma}^{(m)}$;\\
		\@ \@\@ \@ \textit{Step 2}: Optimize $\bm{R}_s^{(m)}$ with (\ref{OptimizeRsSecondType}) and record the optimal value\\
		\@ \@\@ \@ \@ \@\@ \@ \@ \@\@ \@ \@ \@\@ \@ as ${\gamma}^{(m)}$;\\
		\@ \@\@ \@ \textit{Step 3}: $m=m+1$, repeat \textit{step 1} and \textit{step 2} until $\left( \gamma^{(m)} - \tilde{\gamma}^{(m)}\right)/ $\\
		\@ \@\@ \@ \@ \@\@ \@ \@ \@\@ \@ \@ \@\@ \@$\tilde{\gamma}^{(m)}$$\le \varepsilon$; Set $\bm{W}^* = \bm{W}^{(m)}$ and $\bm{R}_s^* = \bm{R}_s^{(m)}$;\\		
		\@ \@\@ \@ \textit{Step 4}: Synthesize transimit signal and receive filter pair. If $\bm{R}_s^*$ or\\
		\@ \@\@ \@ \@ \@\@ \@ \@ \@\@ \@ \@ \@\@ $\bm{W}^*$
		 is rank-one, then $\bm{R}_s^*=\bm{s}^*\left( \bm{s}^*\right) ^{\rm{H}}$ or $\bm{W}^*=\bm{w}^*\left( \bm{w}^*\right) ^{\rm{H}}$;\\
        \@ \@\@ \@ \@ \@\@ \@ \@ \@\@ \@ \@ \@\@  Otherwise, synthesize $\bm{s}^*$ and $\bm{w}^*$ with (\ref{SynthesizeSworstSINR}), (\ref{SynthesizeS}), (\ref{SynthesizeWworstSINR}) and (\ref{SynthesizeW}).\\
		\textbf{Output}: 
		\@ \@\@ \@ ${{\bm{s}^*}}$ and ${{\bm{w}^*}}$\\
		\hline
	\end{tabular}
\end{table}
\subsection{Further discussions on DMDSDR}
In this subsection, the convergences and the computational complexities of DMDSDR are discussed.

First, we discuss the convergences of DMSDR. Without loss of generality, we take the \textit{m}th iteration for example. Let ${{p}}_{\bm{R_s}}^{(m)}$ denote the optimal value for $\tilde{\cal{P}}_{\bm{R}_s}^{(m)}$ with the fixed $\bm{W}^{(m-1)}$, and  ($\bm{\mu}_{\bm{R}_s}^{(m)},\bm{h}_{\bm{R}_s}^{(m)},\bm{Z}_{\bm{R}_s}^{(m)},{\bm{R}_s^{(m)}}$) denote the corresponding optimal solution. Similarly, let ${{p}}_{\bm{W}}^{(m)}$ denote the optimal value for $\tilde{\cal{P}}_{\bm{W}}^{(m)}$ with the fixed $\bm{R}_{s}^{(m)}$, and  ($\bm{\mu}_{\bm{W}}^{(m)},\bm{h}_{\bm{W}}^{(m)},\bm{Z}_{\bm{W}}^{(m)},{\bm{W}}^{(m)}$) denote the corresponding optimal solution. 

Note that  ($\bm{\mu}_{\bm{R}_s}^{(m)},\bm{h}_{\bm{R}_s}^{(m)},\bm{Z}_{\bm{R}_s}^{(m)},{\bm{W}^{(m-1)}}$) is also a feasible point for $\tilde{\cal{P}}_{\bm{W}}^{(m)}$, thus
\begin{equation}
{{p}}_{\bm{R_s}}^{(m)} \le {{p}}_{\bm{W}}^{(m)}.
\label{IncreaseInequalityW}
\end{equation} 
Moreover, ($\bm{\mu}_{\bm{W}}^{(m)},\bm{h}_{\bm{W}}^{(m)},\bm{Z}_{\bm{W}}^{(m)},{\bm{R}}_{s}^{(m)}$) is also a feasible point for $\tilde{\cal{P}}_{\bm{R}_s}^{(m)}$, thus   
\begin{equation}
{{p}}_{\bm{W}}^{(m)}  \le {{p}}_{\bm{R_s}}^{(m+1)}.
\label{IncreaseInequalityRs}
\end{equation}
 
For any $m$, we get the following inequality based on (\ref{StrongDualitEquality}) and (\ref{RelaxtionInequality})
\begin{equation}
\begin{array}{l} {{p}}_{\bm{R_s}}^{(m)} \le 
\mathop {\max }\limits_{{{\bm{R}}_s},{\bm{W}}} \mathop {{\rm{min}}}\limits_{{\bm{\eta}(k)} \le \left| {{\bm{u}(k)}} \right| \le {\bm{\xi}(k)}} \frac{{{{\bm{u}}^{\rm{H}}}{\bm{G}}({{\bm{R}}_{\bm{s}}}{\bm{,W}}){\bm{u}}}}{{{\rm{tr}}\left( {{\bm{W}}{{\bm{R}}_n}} \right)}}\\
\quad  \quad  \le \mathop {\max }\limits_{{{\bm{R}}_s},{\bm{W}}} \frac{{{{{\bm{\hat u}}}^{\rm{H}}}{\bm{G}}({{\bm{R}}_{\bm{s}}}{\bm{,W}}){\bm{\hat u}}}}{{{\rm{tr}}\left( {{\bm{W}}{{\bm{R}}_n}} \right)}}\\
\quad  \quad = \mathop {\max }\limits_{{{\bm{R}}_s},{\bm{W}}} \frac{{{\rm{tr}}({\bm{WA}}{{\bm{R}}_s}{{\bm{A}}^{\rm{H}}})}}{{{\rm{tr}}\left( {{\bm{W}}{{\bm{R}}_n}} \right)}}
\end{array}
\end{equation}
where $\bm{\hat{u}}$ can be any feasible points of $\tilde{\cal{P}}_{1,1}$, and 
${\bm{A}} = \sum\limits_{k = 1}^{K + 1} {\bm{\hat{u}}{(k)}{{{\bm{A}}}_{k - 1}}}$.

Based on the von Neumann's trace theorem\cite{HornMatrix}, we have ${\rm{tr}}\left( {{\bm{AB}}} \right) \le {\rm{tr}}\left( {\bm{A}} \right){\rm{tr}}\left( {\bm{B}} \right)$, for any $\bm{A},\bm{B} \succeq 0$. Thus, the following inequality holds 
\begin{equation}
\begin{array}{l}
\mathop {\max }\limits_{{{\bm{R}}_s},{\bm{W}}} \frac{{{\rm{tr}}({\bm{WA}}{{\bm{R}}_s}{{\bm{A}}^{\rm{H}}})}}{{{\rm{tr}}\left( {{\bm{W}}{{\bm{R}}_n}} \right)}} \le \mathop {\max }\limits_{{{\bm{R}}_s},{\bm{W}}} \frac{{{\rm{tr}}\left( {\bm{W}} \right){\rm{tr}}({\bm{A}}{{\bm{A}}^{\rm{H}}}){\rm{tr}}\left( {{{\bm{R}}_s}} \right)}}{{\lambda _{\min }^{{{\bm{R}}_n}}{\rm{tr}}\left( {\bm{W}} \right)}}\\
\quad \quad \quad \quad \quad \quad \quad \quad = \frac{{{N_T}L{\rm{tr}}({\bm{A}}{{\bm{A}}^{\rm{H}}})}}{{\lambda _{\min }^{{{\bm{R}}_n}}}}
\end{array}
\label{UpBoundInequality}
\end{equation}
where $\lambda _{\min }^{{{\bf{R}}_n}}$ be the minimum eigenvalue of $\bm{R}_{n}$.

Combining (\ref{IncreaseInequalityW}), (\ref{IncreaseInequalityRs}) and (\ref{UpBoundInequality}), we draw the conclusion that the output SINR calculated by DMDSDR monotonically increases with respect to $m$ and is bounded by $\frac{{{N_T}L{\rm{tr}}({\bm{A}}{{\bm{A}}^{\rm{H}}})}}{{\lambda _{\min }^{{{\bm{R}}_n}}}}$. So the proposed MDMSDR converges to a statistical point.

As to the computational complexities, it requires at most $O\left(\left( LN_R\right)^{6.5} + \left( LN_R\right)^{4}K^{2.5} + 
K^{4.5}\right)$ operations to solve ${\tilde {\cal P}_{\bm{W}}^{(m)\prime}}$ and $O\left(\left( LN_T\right)^{6.5} + \left( LN_T\right)^{4}K^{2.5} + 
K^{4.5}\right)$ operations to solve $\tilde{\cal{P}}_{\bm{R}_s}^{(m)}$ at each iteration\cite{NemirovskiLectures}. Moreover, $O\left(\left( LN_T\right)^{2} \right)$ + $O\left(K^{4.5} \right)$ operations are needed to synthesize $\bm{s}^*$, i.e., $O\left(\left( LN_T\right)^{2} \right)$ operations are needed for randomization and $O\left(K^{4.5} \right)$ operations are needed to solve (\ref{SynthesizeSworstSINR})\cite{LuoSemidefinite}. Similarly, $O\left(\left( LN_R\right)^{2} \right)$ + $O\left(K^{4.5} \right)$ are also needed to synthesize $\bm{w}^*$. Note that, in most practical situations, the number $(\left( LN_T\right)^{6.5}$ or $(\left( LN_R\right)^{6.5}$ takes the dominance. To this end, the total computational complexities of DMDSDR are given by $O\left({\rm{max}}\left\lbrace\left( LN_T\right)^{6.5},\left( LN_R\right)^{6.5}\right\rbrace\right)$.

\section{Numerical experiments}
In this section, several numerical experiments are carried out to show the performance of the designed waveform-filter pair. A MIMO radar with $N_T = 4$ transmit antennas and $N_R = 4$ receive antennas is considered. The antenna array is linear and uniform spaced, where the inter-element space is wavelength for transmit antennas and half-wavelength for receive antennas. The carrier frequency is 3GHz, and the code length is $L = 16$ with a sample rate $f_s$=1.5MHz.Meanwhile, the target T is assumed at $\theta_{0} = $30$^\circ$. As to the noise, we assume that noises are correlated in each channel, but independent for different channels. Additionally, it obeys the complex Gaussian distribution ${\bm{n}}\sim{\cal{CN}}\left( {0,{{\bm{R}}_n}} \right)$ with ${{\bm{R}}_n} = {{\bf{I}}_{{N_R}}} \otimes {{\bm{\tilde R}}_n}$,where ${{\bm{\tilde R}}_n}(m,n) = {\sigma^2\beta^{\left| {m - n} \right|}}$, $\sigma^2$ and $\beta$ denote the noise power and correlation coefficient, respectively. Unless specially otherwise specified, $\sigma^2=10$ and $\beta=0.8$ in the following numerical experiments.   

\subsection{Experiments for the spherical uncertainty set}
In this subsection, we consider the spherical uncertainty set. Some experiments are carried out to test the proposed DMSDR algorithms. Note that the algorithms proposed in \cite{chen2009mimo} lack of ability to deal with the CMC waveform, thus, we give a comparison with the algorithms proposed in \cite{karbasi2015robust}.  

In our first experiment, 2 strong scatterers are supposed to exist in the scenario, which means the multipath number is $K=3$(see Fig.\ref{Scenario} and (\ref{MPTargetReceiveEqReciprocity})). The azimuths of multipath  are assumed at $\theta_{1} = -10^\circ$ and $\theta_{2} = -30^\circ$ with the fast time delay $l_1 = 7$ and $l_2 = 5$ sample numbers, respectively. Moreover, the sphere center point is set to be ${{\bm{u}}_0} = {[0.8,0.6{e^{j\pi /3}},0.2{e^{ - j\pi /6}}]^{\rm{T}}}$. 100 random vectors are drawn to synthesize the transmit waveform and receive filter pair ($\bm{s}^*$,$\bm{w}^*$) from $\bm{R}_s^*$. 

Fig.\ref{PerformanceOfWaveformFilterPair} gives performance comparison of the designed waveform-filter pair between DMSDR and algorithms in \cite{karbasi2015robust}(100 samples are randomly picked from the surface of the spherical uncertainty set to construct $\Theta$) with $r=0.5$. In particular, Fig.\ref{PerformanceOfWaveformFilterPair}(a) depicts the transmit-filter antenna pattern calculated by $P(\theta ) = \left| {{{\left( {{{\bm{w}}^*}} \right)}^{\rm{H}}}\left( {{\bm{J}}_0^{\rm{T}} \otimes \left( {{\bm{b}}\left( \theta  \right){{\bm{a}}^{\rm{T}}}\left( \theta  \right)} \right)} \right){{\bm{s}}^*}} \right|$, where azimuths of target and multipath are marked by red and green dotted lines, respectively. An inspection of Fig.\ref{PerformanceOfWaveformFilterPair}(a) reveals that the antenna patterns form peaks near the direction of target and multipath to collect the energy from space. However, the peaks don't completely overlap with these directions due to the coupling term ${\bm{b}}({\theta _0}){{\bm{a}}^{\rm{T}}}({\theta _k}) + {\bm{b}}({\theta _k}){{\bm{a}}^{\rm{T}}}({\theta _0})$ in $\bm{y}_k$. Another phenomenon is that the antenna pattern formed by the two algorithms are almost the same exception for some shaper notches in the antenna pattern formed by DMSDR, which are marked by black ellipses. Fig.\ref{PerformanceOfWaveformFilterPair}(b) outlines the actually output SINR with ($\bm{s}^*$,$\bm{w}^*$) designed by DMSDR and algorithms in \cite{karbasi2015robust} for 50 samples $\bm{u}_i$ which are randomly selected from the uncertainty set with $r = 0.5$. One can see that the actually output SINR for each sample is almost the same for both algorithms. Additionally, Fig.\ref{WorstCaseVersusR3Multipath} depicts the worst-case SINR versus different $r$ with DMSDR and the algorithms in \cite{karbasi2015robust}. Note that the worst-case SINR decreases with respect to the increasing $r$ due to the expansion of uncertainty set, which results in lower worst-case SINR. Compared with the DMSDR algorithm, the worst-case SINR calculated by the algorithms in \cite{karbasi2015robust} fluctuates more seriously when $r$ is relatively large. Given that the surface area of sphere in $n$-dimension is proportion to $r^{n-1}$, this reason can be explained based on the core idea in \cite{karbasi2015robust} that random samples from the surface are used to approximate the uncertainty set. However, the surface area of the sphere increases rapidly with respect to $r$, which means the reduction of sampling density and insufficient approximation accuracy. Finally, combining Fig.\ref{PerformanceOfWaveformFilterPair}(b) and Fig.\ref{WorstCaseVersusR3Multipath} at $r=0.5$, we find that the actually output SINR is significantly higher than the worst-case SINR(about 10.4dB), which demonstrates the effectiveness of the proposed DMSDR algorithm.  
  
Our second experiment is carried out to show the superiority of DMSDR over the algorithms in \cite{karbasi2015robust}, from which one can see the SINR losses owing to insufficient sampling. In this experiment, the number of multipath is increased to $K=$25 and the uncertainty radius $r$ is set to be 0.8 with other parameters being the same as the previous experiment. The azimuth and delay number of multipath returns are generated from the uniform distribution ${\cal U}\left( { - \pi /2, - \pi /2} \right)$ and the uniform integer distribution ${\cal I}\left( { 1, 7} \right)$, respectively. Moreover, the sphere center point is set to be ${\bm{u}_0} = \frac{{{e^{j\pi /4}}}}{{\sqrt {K + 1} }}{{\bf{1}}_{K + 1}}$\vspace{2pt}.

Similarly, Fig.\ref{PerformanceOfWaveformFilterPair25} gives performance comparison of the designed waveform-filter pair between DMSDR and algorithms in \cite{karbasi2015robust} with $r=0.8$. Fig.\ref{PerformanceOfWaveformFilterPair25}(a) depicts the corresponding transmit-receive antenna pattern with azimuths of target and multipath marked by red and green dotted lines, respectively. One can observe that the antenna pattern fails to capture the whole incident energy from different directions due to the lack of enough transmit and receive freedom. As a consequence, the antenna pattern forms a few notches at some multipath directions. Moreover, the two algorithms forms similar antenna patterns expect that DMSDR forms some shaper notches. Alternatively, Fig.\ref{PerformanceOfWaveformFilterPair25}(b) outlines the actually output SINR as well as the worst-case SINR for 50 random samples $\bm{u}_i$. Fortunately, even though antenna pattern couldn't capture the whole energy perfectly, the performance of waveform-filter pair is still satisfactory in terms of actually output SINR, i.e., the actually output SINR is significantly higher than the worst-case SINR for both algorithms. However, the SINR loss of the algorithms in \cite{karbasi2015robust} is obvious in Fig.\ref{PerformanceOfWaveformFilterPair25}(b) compared with DMSDR. Generally speaking, the minimum, average and maximum values of actually output SINR for algorithms in \cite{karbasi2015robust} are 18.1dB, 20.5dB and 22.2dB, respectively. Meanwhile, the minimum, average and maximum values of actually output SINR for DMSDR are 18.8dB, 21.1dB and 22.8dB, respectively. The SINR loss is an immediate sequence of insufficient sampling. These results demonstrate the competitiveness of DMSDR compared with its counterpart in\cite{karbasi2015robust}.

\begin{figure}[!t]
	\centering
	\begin{tabular}{c}
		\includegraphics[width=2.4in]{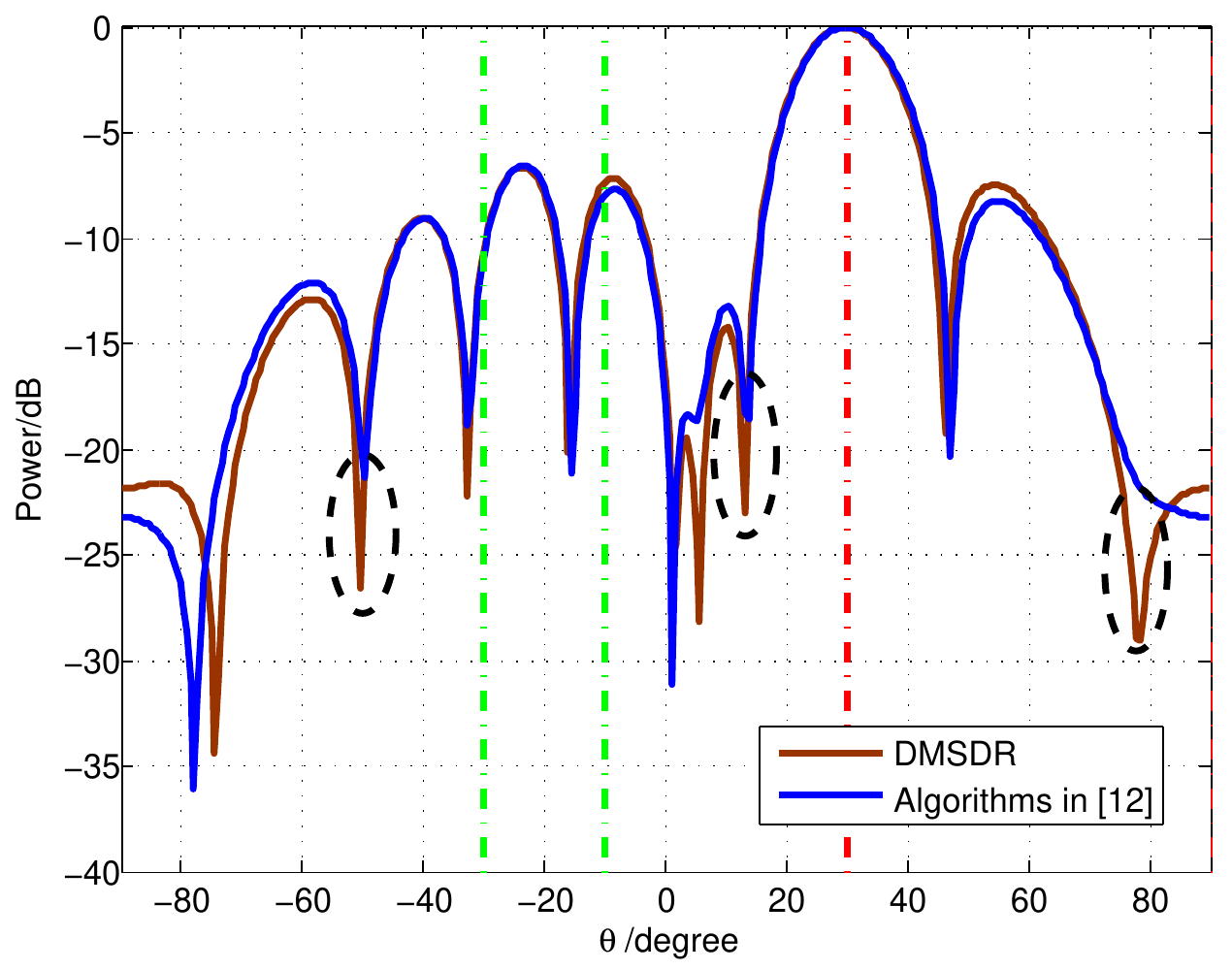}\\
		(a)\\
		\includegraphics[width=2.4in]{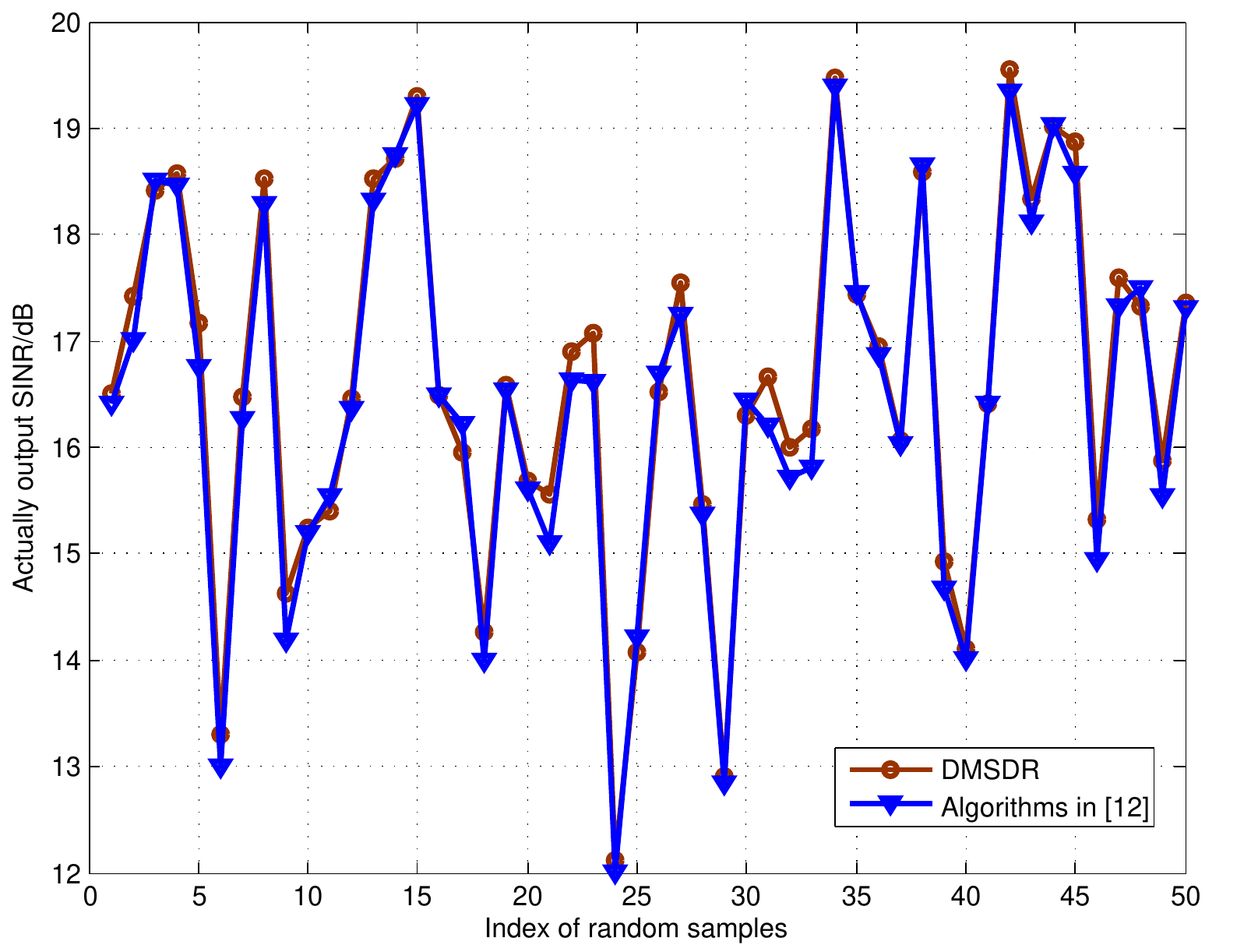}\\
		(b)\\
	\end{tabular}
	\centering
	\caption{Performance of waveform-filter pair with 2 multipath: (a)Transmit-receive antenna pattern. (b) The actually output SINR.}
	\label{PerformanceOfWaveformFilterPair}
\end{figure} 

\begin{figure}[!t]
	\centering
	\includegraphics[width=2.4in]{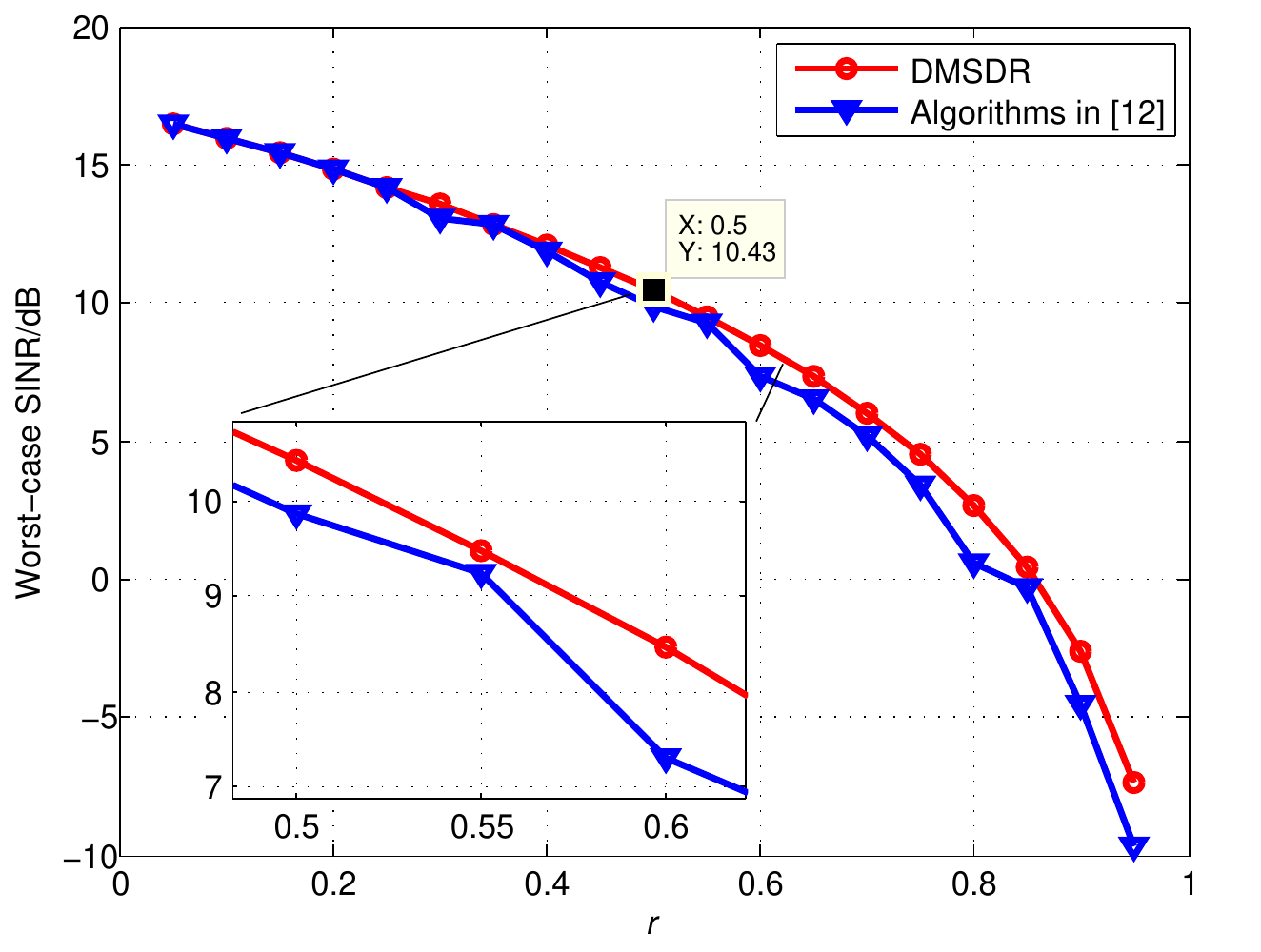}
	\caption{Worst-case SINR versus the uncertainty radius $r$ with 2 multipaths}
	\label{WorstCaseVersusR3Multipath}
\end{figure}

\begin{figure}[!t]
	\centering
	\begin{tabular}{c}
		\includegraphics[width=2.4in]{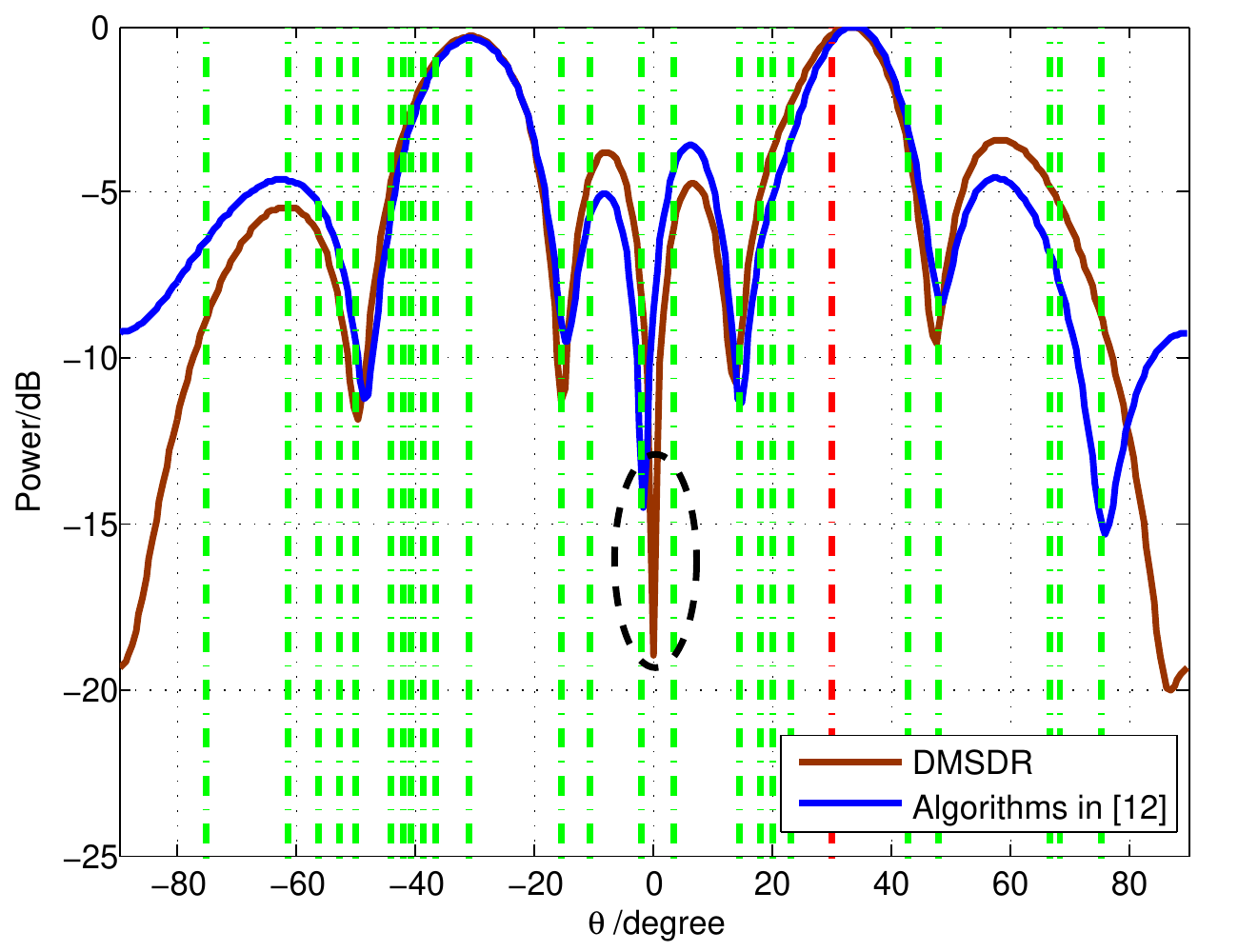}\\
		(a)\\
		\includegraphics[width=2.4in]{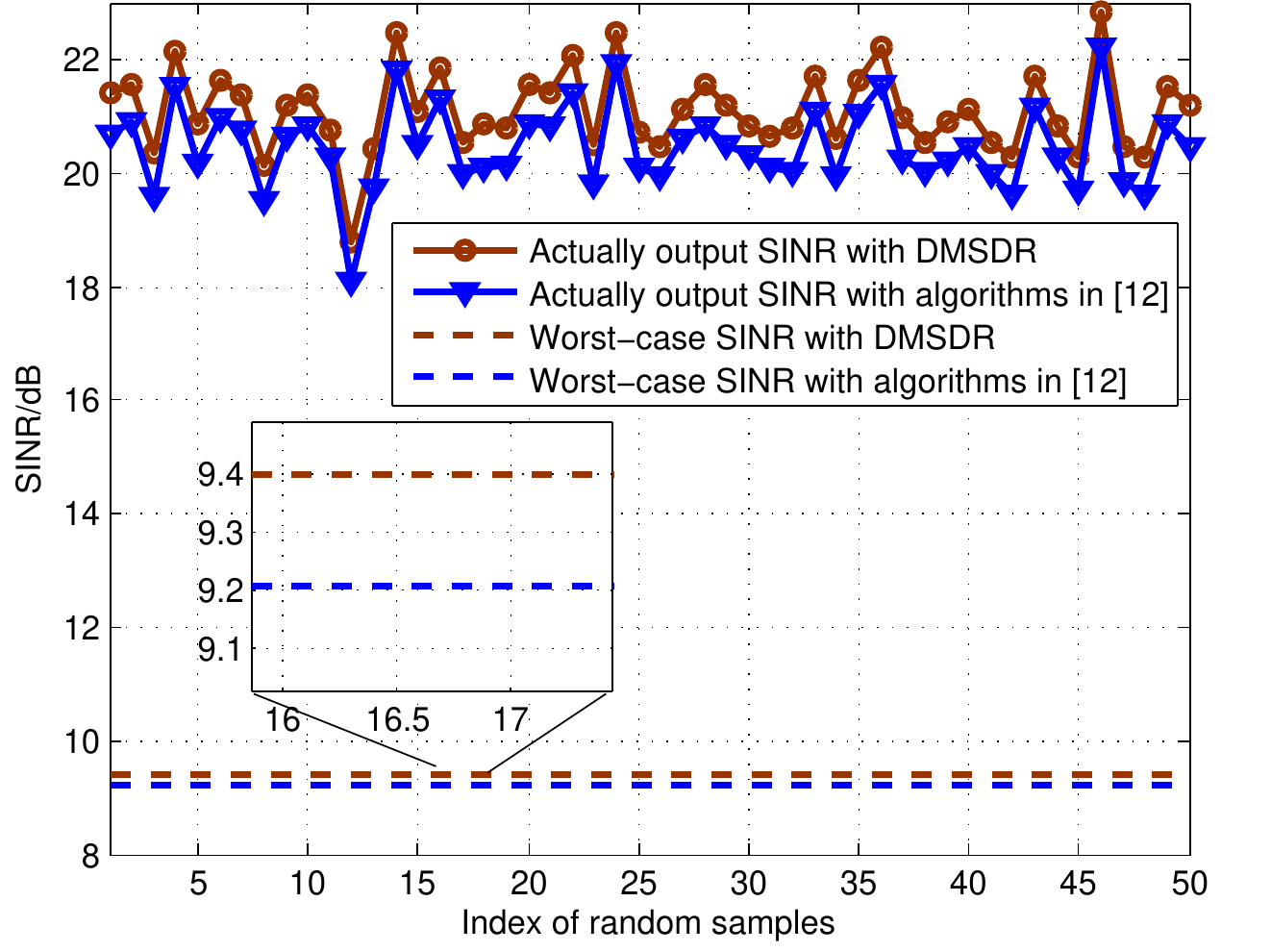}\\
		(b)\\
	\end{tabular}
	\centering
	\caption{Performance of waveform-filter pair with 25 multipath: (a)Transmit-receive antenna pattern. (b) The actually output SINR.}
	\label{PerformanceOfWaveformFilterPair25}
\end{figure}

\subsection{Experiments for the annular uncertainty set}
This subsection is devoted to robust waveform-filter design against the annular uncertainty set with DMDSDR. In the following experiment, the number of multipath is $K=2$. The azimuths of multipath are assumed at $\theta_{1} = -30^\circ$ and $\theta_{2} = -10^\circ$ with the fast time delay $l_1 = 3$ and $l_2 = 7$ samples, respectively.The transmit waveform is initialized by a pseudo random phase coded signal and the tolerable error is set to be $\varepsilon = 0.001$.  

Denoted $\Theta (a,b,c,d)$ by the annular uncertainty set, where $\bm{\eta} = {\left[ {a,b,c} \right]^{\rm{T}}}$ and ${\bm{\xi }}  = {\left[ {a+d,b+d,c+d} \right]^{\rm{T}}}$. Fig.\ref{PerformanceOfWaveformFilterPairAnnular} gives performance of the designed waveform-filter pair by DMDSDR with $\Theta(2,1,0.5,2)$. More exactly, Fig.\ref{PerformanceOfWaveformFilterPairAnnular}(a) depicts the transmit-receive antenna pattern. Instead of forming several peaks to capture the energy from different directions, the antenna pattern just forms a peak at the direction of target, which is very different from the cases under spherical uncertainty. As aforementioned, the annular uncertainty set means no phase information of returns. Therefore, collecting energy from all directions may result in energy cancelling. To this end, the robust way is to collect the strongest energy, which leads to the antenna pattern in Fig.\ref{PerformanceOfWaveformFilterPairAnnular}(a). In order to investigate the output SINR achieved by DMDSDR, 50 samples $\bm{u}_i$ are randomly picked from the annular uncertainty set, where the amplitude and the phase of $\bm{u}_j(k)$ are generated from the uniform distribution ${\cal U}\left( {  \bm{\eta}(k) ,  \bm{\xi}(k)} \right)$ and ${\cal U}\left( { -\pi  , \pi}\right) $, respectively.  Fig.\ref{PerformanceOfWaveformFilterPairAnnular}(b) depicts the actually output SINR for each sample as well as the worst-case SINR. We find that the actually output SINR is higher than the worst-case SINR, which demonstrates the effectiveness of the proposed DMDSDR algorithm.

In order to give an insight into the effects on the worst-case SINR caused by the parameters of the annular uncertainty set, Fig.\ref{SINRCurveVersusDifferentParameter} outlines the worst-case SINR with respect to different parameters versus the number of iterations. As expected, the worst-case SINR monotonically increases with respect to the number of iterations. Comparisons among the SINR curves in Fig.\ref{SINRCurveVersusDifferentParameter} reveals that the worst-case SINR is mainly affected by the parameter $a$ which represents the lower bound amplitude of the strongest path. These in turn verify the antenna pattern properties in Fig.\ref{PerformanceOfWaveformFilterPairAnnular}(b).

Finally, we are going to see an interesting phenomenon. As shown in Fig.\ref{SINRCurveVersusDifferentParameter}, swapping values of $b$ and $c$ has few effects on the worst-case SINR(see $\Theta(2,1,0.5,2)$ and $\Theta(2,0.5,1,2)$ curves in Fig.\ref{SINRCurveVersusDifferentParameter}). Nevertheless, we will illustrate that swapping values of $a$ and $b$(or $a$ and $c$) will affect the worst-case SINR. Before that, we must clarify its physical meaning of $\bm{\eta}(k)>\bm{\eta}(1)$. Note that $\bm{\eta}(1)$ denotes the lower bound amplitude of the target returns from line of sight, while $\bm{\eta}(k)$ denotes the lower bound amplitude of the target returns from the direction of $(k-1)$th multipath. Thus, $\bm{\eta}(k)>\bm{\eta}(1)$ means that a repeater which amplifies the signal exists at the direction $\theta_{k-1}$. Fig.\ref{EffectsCausedBySwapping} shows the effects on worst-case SINR and transmit-receive antenna pattern caused by swapping values of $a$, $b$ and $c$. One can find that the worst-case SINR changes with respect to the permutation of $a$, $b$ and $c$, which can be explained from the antenna pattern(see Fig.\ref{EffectsCausedBySwapping}(b)) whose mainlobe always points at the direction of the strongest returns.

\begin{figure}[!t]
	\centering
	\begin{tabular}{c}
		\includegraphics[width=2.4in]{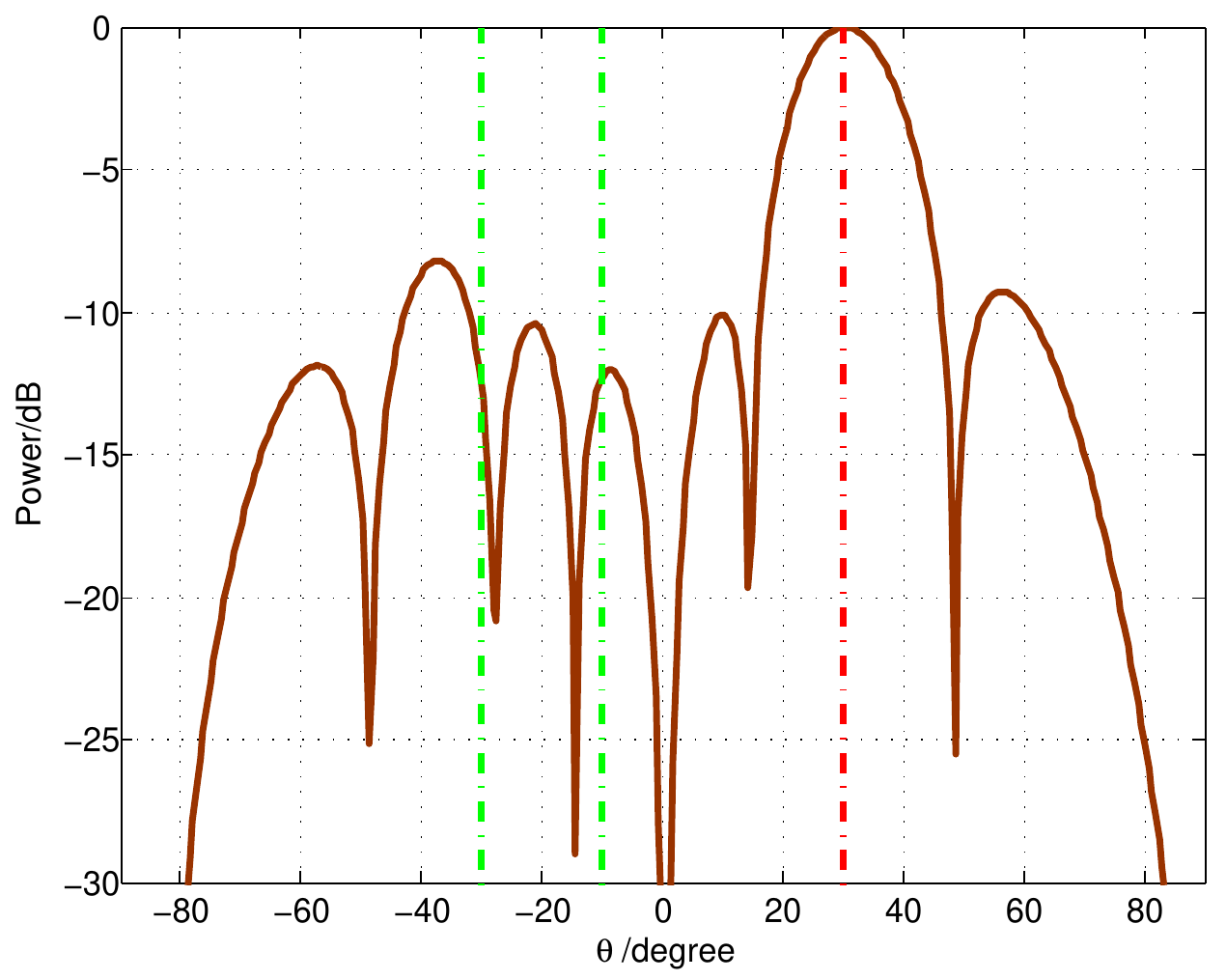}\\
		(a)\\
		\includegraphics[width=2.6in]{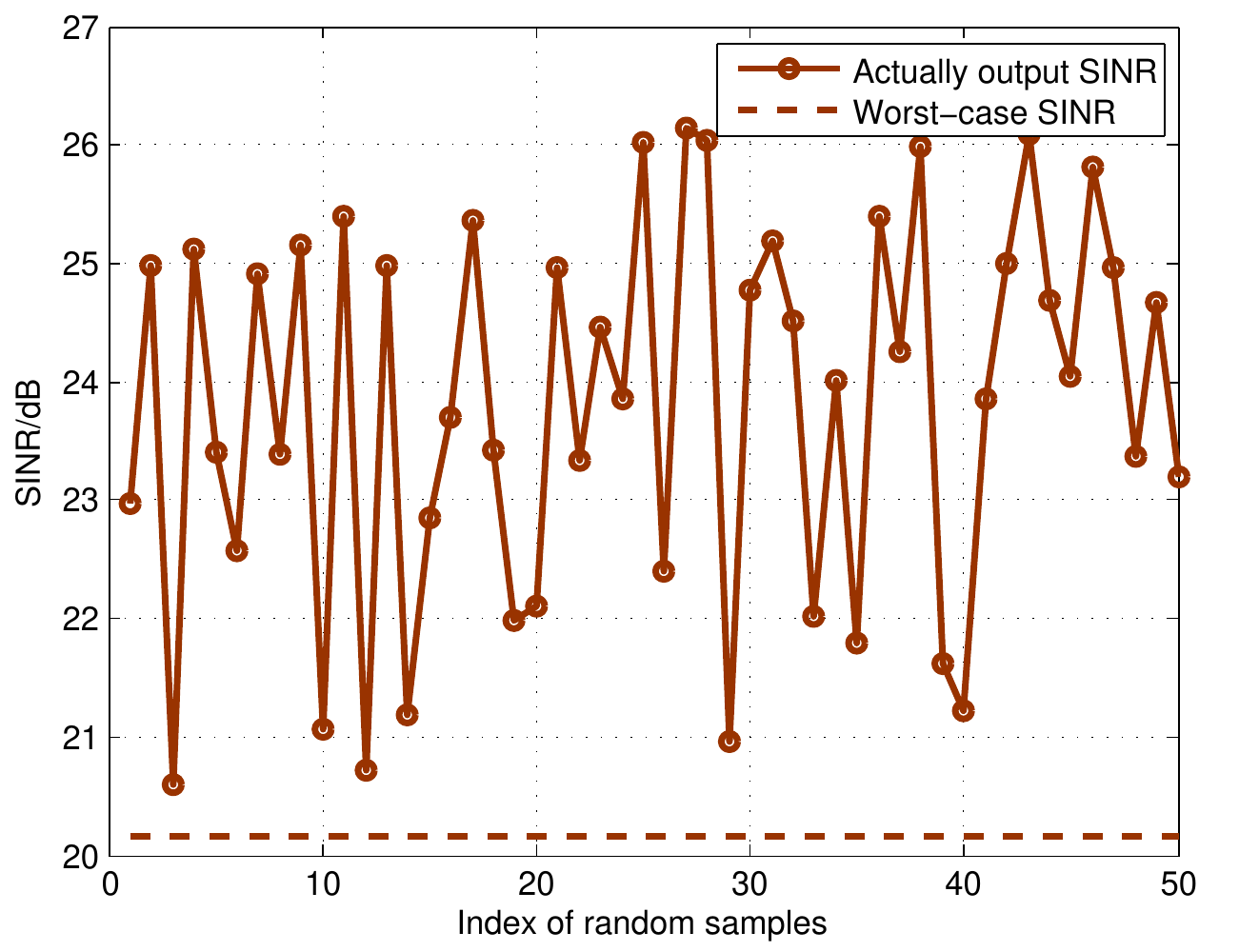}\\
		(b)\\
	\end{tabular}
	\centering
	\caption{Performance of waveform-filter pair with 2 multipath under the annular uncertainty set: (a)Transmit-receive antenna pattern. (b) The actually output SINR.}
	\label{PerformanceOfWaveformFilterPairAnnular}
\end{figure} 

\begin{figure}[!t]
	\centering
	\includegraphics[width=2.4in]{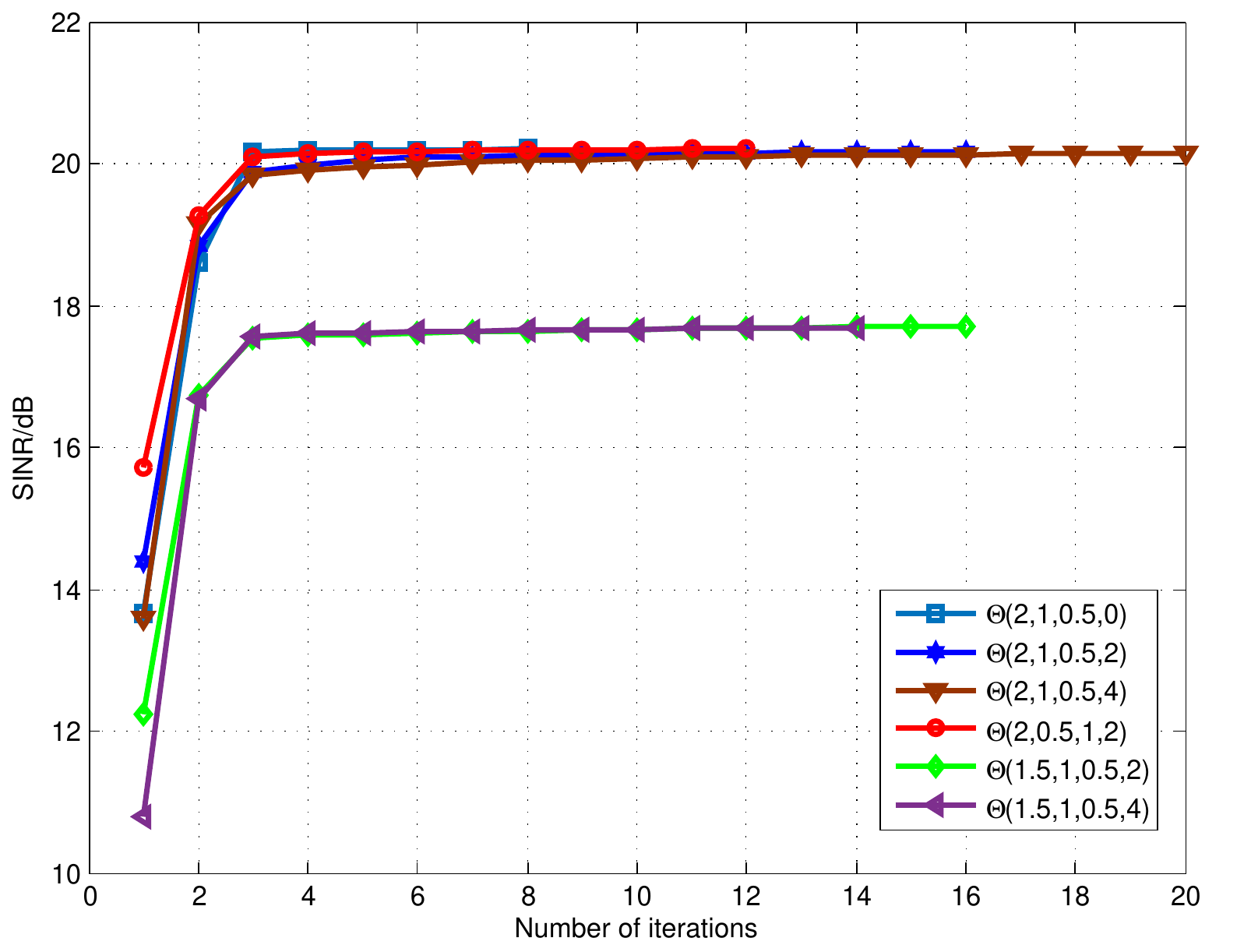}
	\caption{Worst-case SINR versus different annular uncertainty parameters}
	\label{SINRCurveVersusDifferentParameter}
\end{figure}

\begin{figure}[!t]
	\centering
	\begin{tabular}{c}
		\includegraphics[width=2.4in]{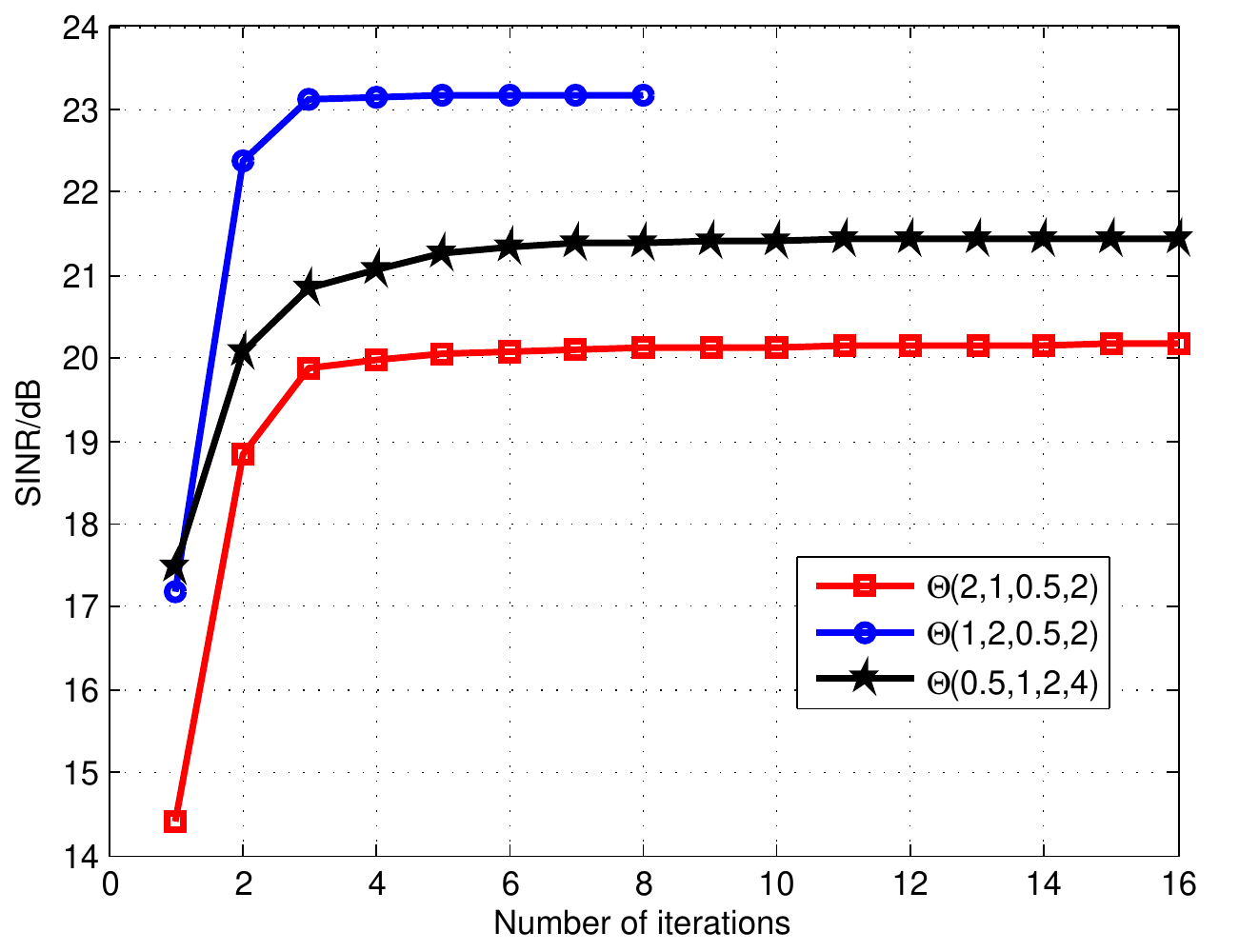}\\
		(a)\\
		\includegraphics[width=2.3in]{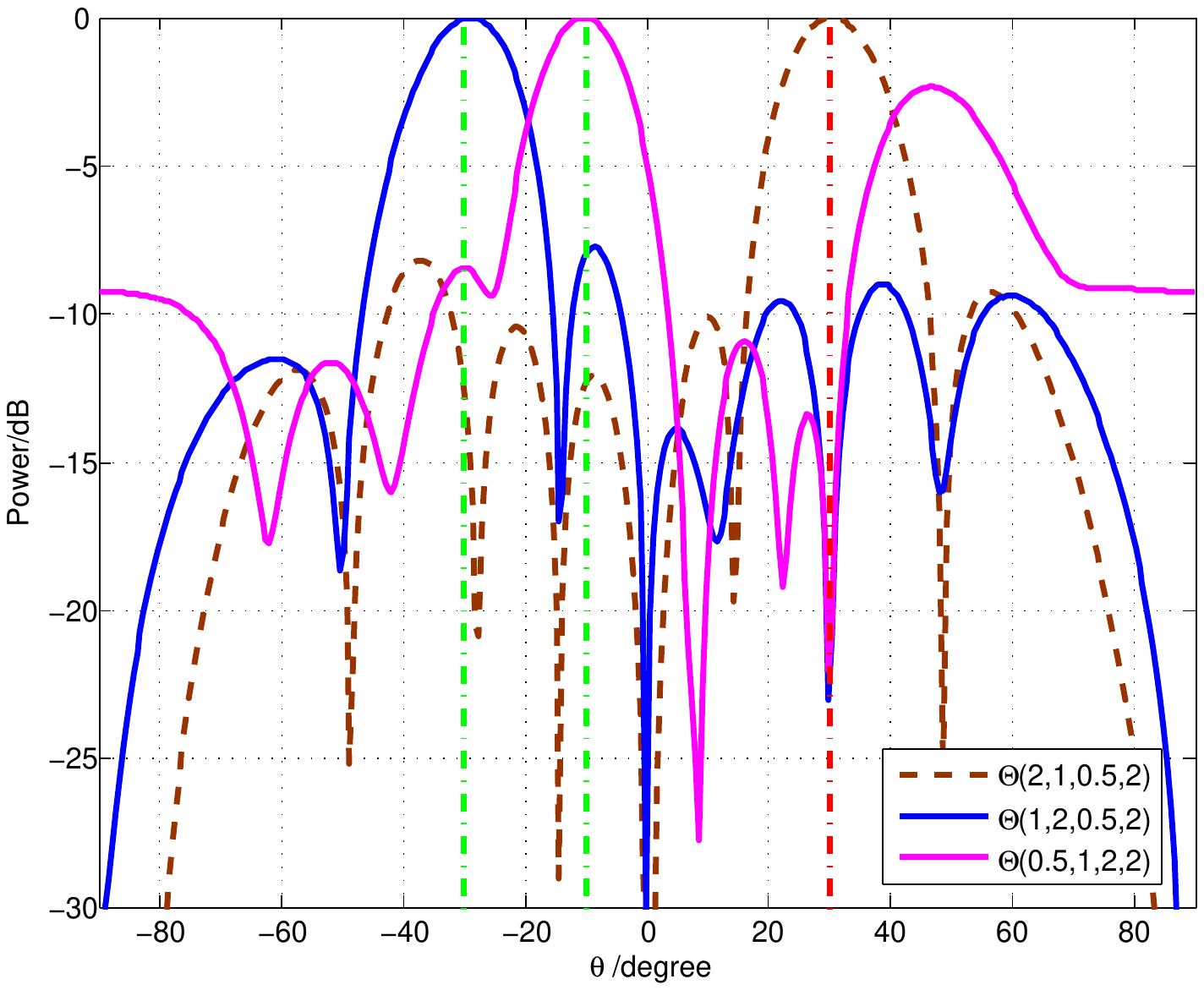}\\
		(b)\\
	\end{tabular}
	\centering
	\caption{Effects caused by swapping parameters: (a)Worst-case SINR iteration curve. (b)Transmit-receive antenna pattern.}
	\label{EffectsCausedBySwapping}
\end{figure}

%
%
%

\section{Conclusion}
In this paper, robust design problems of waveform-filter for extended target detection in the presence of multipath with the MIMO radar are considered. In order to deal with the imprecise prior knowledge of the target and multipath scattering coefficients,  the worst-case SINR is used as the designing criterion. Two different types of the uncertainty sets are studied. The first one is the spherical uncertainty set, which means that the actual scattering coefficients belong to a scaled ball centered around an a priori known scattering coefficients. The second one is the annular uncertainty set, which means that the amplitude of scattering coefficients can be roughly estimated in advance, but the phase information can't be obtained.

For the spherical uncertainty set, we propose the DMSDR algorithm to solve this problem. The Lagrange duality function is utilized to convert the inner minimization problem to a maximization problem. Then, the maximization problem is approximate by a convex problem with the SDR, which can be solved in polynomial time.

For the annular uncertainty set, we propose the DMDSDR algorithm to solve this problem. Note that the annular uncertainty set is non-convex. The SDR method is used to approximate the inner minimization problem with a SDP problem, then, it is be converted to a maximization problem based on Lagrange duality. We devise a cyclic SDR method to optimize the covariance of transmit waveform and receive filter alternatively. Additionally, the convergences of DMDSDR are proved theoretically.

At the analysis stage, some numerical experiments are presented. It can be observed that both the DMSDR algorithm and DMDSDR algorithm provide the relatively stable output SINR, which highlights the robustness of the designed transmit waveform and receive filter pair. Another interesting result is that the optimal waveform-filter pair against the spherical uncertainty set attempts to capture the energy from all directions, while the optimal waveform-filter pair against the annular uncertainty set always tracks the energy from the direction of the strongest returns.

Our future researches may include the waveform design under more practical constraints, such as PAR constraint, similarity constraint and spectrally compatible constraint.  Moreover, waveform design for other purposes in the presence of multipath, for instance, location and recognition, will be also interesting.

\appendices
\section{Proof of Proposition 1}
For a given $\bm{s}_0$, the optimal objective in ${{\cal P}_1}$ and ${{\cal P}'_1}$ can be represented as $\mathop {{\rm{ max}}}\limits_{\bm{w}} \mathop {{\rm{min}}}\limits_{{\bm{u}}}{\rm{ }}\frac{{{{\left| {{{\bm{w}}^{\rm{H}}}{\bm{Yu}}} \right|}^2}}}{{{{\bm{w}}^{\rm{H}}}{{\bm{R}}_n}{\bm{w}}}}$ and $\mathop {{\rm{ min}}}\limits_{\bm{u}} \mathop {{\rm{max}}}\limits_{{\bm{w}}}{\rm{ }}\frac{{{{\left| {{{\bm{w}}^{\rm{H}}}{\bm{Yu}}} \right|}^2}}}{{{{\bm{w}}^{\rm{H}}}{{\bm{R}}_n}{\bm{w}}}}$, respectively.

Note that \vspace{2pt} $\frac{{{{\left| {{{\bm{w}}^{\rm{H}}}{\bm{Yu}}} \right|}^2}}}{{{{\bm{w}}^{\rm{H}}}{{\bm{R}}_n}{\bm{w}}}} = \frac{{{{\bm{u}}^{\rm{H}}}{{\bm{Y}}^{\rm{H}}}{\bm{w}}{{\bm{w}}^{\rm{H}}}{\bm{Yu}}}}{{{{\bm{w}}^{\rm{H}}}{{\bm{R}}_n}{\bm{w}}}}$, is a convex with respect to $\bm{u}$, and the constraint ${\left\| {{\bm{u}} - {{\bm{u}}_0}} \right\|_2} \le r$ is also a convex set. Therefore, according to Theorem 1 in \cite{kim2006robust}, we get 
\begin{equation}
 \mathop {{\rm{ max}}}\limits_{\bm{w}} \mathop {{\rm{min}}}\limits_{{\bm{u}}}{\rm{ }}\frac{{{{\left| {{{\bm{w}}^{\rm{H}}}{\bm{Yu}}} \right|}^2}}}{{{{\bm{w}}^{\rm{H}}}{{\bm{R}}_n}{\bm{w}}}} = \mathop {{\rm{ min}}}\limits_{\bm{u}} \mathop {{\rm{max}}}\limits_{{\bm{w}}}{\rm{ }}\frac{{{{\left| {{{\bm{w}}^{\rm{H}}}{\bm{Yu}}} \right|}^2}}}{{{{\bm{w}}^{\rm{H}}}{{\bm{R}}_n}{\bm{w}}}},
\end{equation}
with the constraint ${\left\| {{\bf{u}} - {{\bf{u}}_0}} \right\|_2} \le r$.

Consequently, we get 
\begin{equation}
\mathop {{\rm{ max}}}\limits_{\bm{w},\bm{s}} \mathop {{\rm{min}}}\limits_{{\bm{u}}}{\rm{ }}\frac{{{{\left| {{{\bm{w}}^{\rm{H}}}{\bm{Yu}}} \right|}^2}}}{{{{\bm{w}}^{\rm{H}}}{{\bm{R}}_n}{\bm{w}}}} = \mathop {{\rm{ max}}}\limits_{\bm{s}} \mathop {{\rm{ min}}}\limits_{\bm{u}} \mathop {{\rm{max}}}\limits_{{\bm{w}}}{\rm{ }}\frac{{{{\left| {{{\bm{w}}^{\rm{H}}}{\bm{Yu}}} \right|}^2}}}{{{{\bm{w}}^{\rm{H}}}{{\bm{R}}_n}{\bm{w}}}},
\end{equation}
with the constraints ${\left\| {{\bf{u}} - {{\bf{u}}_0}} \right\|_2} \le r$ and $\left| {{\bm{s}(i)}} \right| = 1$.

Thus, Thus we complete the proof of Proposition 1.

\section{Proof of Proposition 2}
The Lagrange function of (\ref{InnerMinimizationFirstType}) can be represented as 
\begin{equation*}
{\cal L}(\mu, \bm{u} ){\rm{ =  }} {\rm{ }}{{\bm{u}}^{\rm{H}}}{{\bm{Y}}^{\rm{H}}}{\bm{R}}_n^{ - 1}{\bm{Yu}} + \mu \left( {\left\| {{\bm{u}} - {{\bm{u}}_0}} \right\|_2^2 - {r^2}} \right),
\end{equation*}
and 
\begin{equation*}
\frac{{\partial {\cal L}}}{{\partial {\bm{\bar u}}}}{\rm{ =  }}{{\bm{Y}}^{\rm{H}}}{\bm{R}}_n^{ - 1}{\bm{Yu}} + \mu {\bm{u}} - \mu {{\bm{u}}_0},
\end{equation*}
Let $\frac{{\partial {\cal L}}}{{\partial {\bm{\bar u}}}}{\rm{ =  }}0$, then we get ${\bm{u}} = \mu {\left( {{{\bm{Y}}^{\rm{H}}}{\bm{R}}_n^{ - 1}{\bm{Y}} + \mu \bf{I}} \right)^{ - 1}}{{\bm{u}}_0}$. Thus, the Lagrange dual function can be represented as
\begin{equation*}
g\left( \mu  \right){\rm{ = }} - {\mu ^2}{\bm{u}}_0^{\rm{H}}{\left( {{{\bm{Y}}^{\rm{H}}}{\bm{R}}_n^{ - 1}{\bm{Y}} + \mu \bf{I}} \right)^{ - 1}}{\bm{u}}_0^{} + \mu \left( {{\bm{u}}_0^{\rm{H}}{{\bm{u}}_0} - {r^2}} \right),
\end{equation*}
and the dual problem is 
\begin{equation*}
\begin{array}{l}
\mathop {{\rm{max}}}\limits_\mu  {\rm{ }}g\left( \mu  \right), \quad \quad 
s.t.\quad \mu  \ge 0
\end{array}.
\end{equation*}
Thus we complete the proof of Proposition 2.

\section{Proof of Proposition 3}

\begin{equation*}
\begin{array}{l}
{\bm{T}}({\bm{s}}) = {{\bm{Y}}^{\rm{H}}}{\bm{R}}_n^{ - 1}{\bm{Y}} \vspace{3pt}\\
\quad \quad \ = \left[ \begin{array}{l}
{{\bm{s}}^{\rm{H}}}{\bm{A}}_0^{\rm{H}}\\
{{\bm{s}}^{\rm{H}}}{ {\bm{A}}}_1^{\rm{H}}\\
\quad  \vdots \\
{{\bm{s}}^{\rm{H}}}{{\bm{A}}}_K^{\rm{H}}
\end{array} \right]{\bm{R}}_n^{ - 1}\left[ {{{\bm{A}}_0}s,{{{{\bm{ A}}}}_1}{\bm{s}},...{{{ {\bm{A}}}}_K}{\bm{s}}} \right]\vspace{3pt}\\
\quad \quad \ = \left[ {\begin{array}{*{20}{c}}
	{{{\bm{s}}^{\rm{H}}}{\bm{A}}_0^{\rm{H}}{\bm{R}}_n^{ - 1}{{\bm{A}}_0}{\bm{s}}}& \cdots &{{{\bm{s}}^{\rm{H}}}{\bm{A}}_0^{\rm{H}}{\bf{R}}_n^{ - 1}{{{ {\bm{A}}}}_K}{\bf{s}}}\\
	\vdots & \ddots & \vdots \\
	{{{\bm{s}}^{\rm{H}}}{ {\bm{A}}}_K^{\rm{H}}{\bm{R}}_n^{ - 1}{{\bm{A}}_0}{\bm{s}}}& \cdots &{{{\bm{s}}^{\rm{H}}}{ {\bm{A}}}_K^{\rm{H}}{\bm{R}}_n^{ - 1}{{{ {\bm{A}}}}_K}{\bm{s}}}
	\end{array}} \right],
\end{array}
\end{equation*}
and 
\begin{equation*}
\begin{array}{l}
{\bm{T}}\left(i,j \right) =  {{{\bm{s}}^{\rm{H}}}{ {\bm{A}}}_{i-1}^{\rm{H}}{\bm{R}}_n^{ - 1}{{\bm{A}}_{j-1}}{\bm{s}}}\\
={\rm{tr}}\left( {{\bm{A}}_{i - 1}^{\rm{H}}{\bm{R}}_n^{ - 1}{\tilde {\bm{A}}}_{j - 1}{{\bm{R}}_s}} \right)
\end{array}.
\end{equation*}

Thus we complete the proof of Proposition 3.

\section{Proof of Proposition 4}

Let $\bm{E}_k = {\bm{e}_k}{\bm{e}_k^{\rm{H}}}$, we reformulate  $\tilde{\cal{P}}_{1,2}$ as 

\begin{equation}
\left\{ \begin{array}{l}
\mathop {{\rm{min}}}\limits_{{{\bm{R}}_u}} {\rm{ tr}}\left( {{{\bm{Y}}^{\rm{H}}}{\bm{w}}{\bm{w}^{\rm{H}}}{\bm{Y}}{{\bm{R}}_u}} \right)\\
s.t.\quad {{\bm{\eta}(k)}^2} - {\rm{tr}}\left( {{{\bm{E}}_k}{{\bm{R}}_u}} \right) \le 0\\
\quad \quad \;{\rm{tr}}\left( {{{\bm{E}}_k}{{\bm{R}}_u}} \right) - { {\bm{\xi}(k)}^2} \le 0\\
\quad \quad {{\bm{R}}_u} \succeq 0
\end{array} \right..
\label{InnerMinimizationSecondTypeSDP}
\end{equation}
And the Lagrange function of (\ref{InnerMinimizationSecondTypeSDP}) can be represented as
\begin{equation}
\begin{array}{l}
{{\cal \bm{L}}}({\bm{\mu }},{\bm{h}},{\bm{Z}},{{\bm{R}}_u}) = {\rm{tr}}\left( {{{\bm{Y}}^{\rm{H}}}{\bm{w}}{\bm{w}^{\rm{H}}}{\bm{Y}}{{\bm{R}}_u}} \right) - {\rm{tr}}({\bm{Z}}{{\bm{R}}_u}) + \\
\sum\limits_{k = 1}^{K + 1} {{\bm{\mu}(k)}\left( {{{\bm{\eta}(k)}^2} - {\rm{tr}}\left( {{{\bm{E}}_k}{{\bm{R}}_u}} \right)} \right) + {\bm{h}(k)}\left( {{\rm{tr}}\left( {{{\bm{E}}_k}{{\bm{R}}_u}} \right) - {{\bm{\xi}(k)}^2}} \right)} 
\end{array}
\end{equation}
where $\bm{\lambda}, \bm{v}, \bm{Z}$ are the corresponding dual variables. And the Lagrange dual function
\begin{equation}
g({\bm{\mu }},{\bm{h}},{\bm{Z}}) = \mathop {\inf }\limits_{{{\bm{R}}_u}} {\rm{ }}{\cal L}({\bm{\mu }},{\bm{h}},{\bm{Z}},{{\bm{R}}_u}).
\end{equation}

Note that ${{\cal \bm{L}}}({\bm{\mu }},{\bm{h}},{\bm{Z}},{{\bm{R}}_u})$ is a linear function with respect to ${{\bm{R}}_u}$, so $g({\bm{\mu }},{\bm{h}},{\bm{Z}})$ is bounded if and only if ${{\bm{Y}}^{\rm{H}}}{\bm{w}}{\bm{w}^{\rm{H}}}{\bm{Y}} - {\bm{Z}} + \sum\limits_{k = 1}^{K + 1} {\left( {{\bm{h}(k)}-{\bm{\mu}(k)}} \right){{\bf{E}}_k}}  = {\bf{0}}$, in which case $g({\bm{\mu }},{\bm{h}},{\bm{Z}}) = \sum\limits_{k = 1}^{K + 1} {{\bm{\mu}(k)}{{\bm{\eta}(k)}^2} - {\bm{h}(k)}{{\bm{\xi}(k)}^2}}$.

Let ${\bm{f}} = {\bm{\eta }} \odot {\bm{\eta }}$, ${\bm{g}} = {\bm{\xi }} \odot {\bm{\xi }}$, and we get the dual problem of $\tilde{\cal{P}}_{1,2}$
\begin{equation}
\left\{ \begin{array}{l}
\mathop {{\rm{max}}}\limits_{{\bm{\mu }},{\bm{h}},{\bm{Z}}} {\rm{ }}{{\bm{u}}^{\rm{T}}}{\bm{f} - }{{\bm{h}}^{\rm{T}}}{\bm{g}}
\\
s.t.\ {{\bm{Y}}^{\rm{H}}}{\bm{w}}{\bm{w}^{\rm{H}}}{\bm{Y}} - {\bm{Z}} + \sum\limits_{k = 1}^{K + 1} {\left( { {\bm{h}(k)}-{\bm{\mu}(k)}} \right){{\bm{E}}_k}}  = {\bf{0}}\\
\quad \quad {{\bm{\mu}(k)}} \ge 0,{\bm{h}(k)} \ge 0, k=1,2,3...,K+1,\\
\quad \quad {\bm{Z}} \succeq 0
\end{array} \right..
\end{equation}
 
Thus we complete the proof of Proposition 4.

\section{Proof of Proposition 5}
Note that the objective of ${\tilde {\cal P}_{\bm{W}}^{(m)\prime}}$ is equivalent to the \vspace{3pt} objective of ${\tilde {\cal P}_{\bm{W}}^{(m)}}$, \vspace{3pt} but the feasible set of ${\tilde {\cal P}_{\bm{W}}^{(m)\prime}}$ is included by the feasible set of ${\tilde {\cal P}_{\bm{W}}^{(m)}}$, thus $p_{\bm{W}}^{(m)\prime} \le p_{\bm{W}}^{(m)}$.\vspace{3pt}

Moreover, for any feasible solutions $(\bm{\mu},\bm{h},\bm{Z},\bm{W})$ \vspace{3pt} for ${\tilde {\cal P}_{\bm{W}}^{(m)}}$, there exists $(\bm{\mu}/\xi,\bm{h}/\xi,\bm{Z}/\xi,\bm{W}/\xi)$ is a \vspace{3pt} feasible solution for ${\tilde {\cal P}_{\bm{W}}^\prime}$ with the same objective value in ${\tilde {\cal P}_{\bm{W}}}$ where $\xi = {{\rm{tr}}(\bm{R}_n{\bm{W}})}$. So,  $p_{\bm{W}}^{(m)\prime} = p_{\bm{W}}^{(m)}$, and the optimal solution $(\bm{\mu}^*,\bm{h}^*,\bm{Z}^*,\bm{W}^{(m)})$ for ${\tilde {\cal P}_{\bm{W}}^{(m)\prime}}$ is also a optimal solution for ${\tilde {\cal P}_{\bm{W}}^{(m)}}$.

Thus we complete the proof of Proposition 5.
          

\bibliographystyle{IEEEtran}
\bibliography{RobustMIMOWaveformDesignExtended20200923}     

\end{document}